\documentclass[raa]{raa}            

\usepackage{graphicx,times}             
\usepackage{natbib}
\usepackage{amssymb,amsmath}
\bibpunct{(}{)}{;}{a}{}{,}

\usepackage{amsmath}	
\usepackage{amssymb}	
\usepackage{wasysym}
\usepackage{newtxtext,newtxmath}
\usepackage{subcaption}
\usepackage{gensymb}

\usepackage{hyperref}
\hypersetup{colorlinks = true, linkcolor = green, anchorcolor = red, citecolor = blue, filecolor = red, pagecolor = red, urlcolor = red}

\begin{document}

   \title{\textit{AstroSat} observations of the Be/X-ray binary XTE J1946+274 during 2018 and 2021 outbursts}

   \volnopage{Vol.0 (20xx) No.0, 000--000}      
   \setcounter{page}{1}          
   \author{Amar Deo Chandra
   \inst{1}$^{,*}$, Jayashree Roy\inst{2}$^{,*}$, P. C. Agrawal\inst{3}$^{,\#}$
   }

   \institute{Indian Institute of Science Education and Research Kolkata, Mohanpur 741246, West Bengal, India; {\it amar.deo.chandra@gmail.com}\\
        \and
             Inter-University Center for Astronomy and Astrophysics, Post Bag 4, Pune 411007, Maharashtra, India; {\it jayashree@iucaa.in}\\
	\and
Department of Astronomy and Astrophysics, Tata Institute of Fundamental Research, Homi Bhabha Road, Mumbai 400005, Maharashtra, India\\
$^*$Corresponding authors.\\
$^\#$Senior professor (retired).\\
\vs \no
   {\small Received 20XX Month Day; accepted 20XX Month Day}
}

\abstract{ We present the timing and spectral studies of the Be/X-ray binary XTE J1946+274 during its 2018 and 2021 giant outbursts using observations with the SXT and the LAXPC instruments on the \textit{AstroSat} satellite. Unlike the 1998 and 2010 outbursts, where a giant outburst was followed by several low intensity periodic outbursts, the 2018 and 2021 outbursts were single outbursts. The X-ray pulsations are detected over a broad energy band covering 0.5-80 keV from the compact object. We construct the  spin evolution history of the pulsar over two decades and find that the pulsar spins-up during the outbursts but switches to spin-down state in the quiescent periods between the outbursts. Energy resolved pulse profiles generated in several bands in 0.5-80 keV show that the pulse shape varies with the energy. The energy spectrum of the pulsar is determined for the 2018 and 2021 outbursts. The best fit spectral models require presence of Cyclotron Resonant Scattering Feature (CRSF) at about 43 keV in the energy spectra of both the outbursts. We find indication of possible reversal
in the correlation between the cyclotron line energy and luminosity which needs to be ascertained from future observations. Using the best fit spectra the X-ray luminosity of XTE J1946+274 is inferred to be $2.7 \times 10^{37}$\,erg\,s$^{-1}$ for the 2018 observations and $2.3 \times 10^{37}$\,erg\,s$^{-1}$ for the 2021 observations. We discuss possible mechanisms which can drive outbursts in this transient Be X-ray binary.}

   \authorrunning{A. D. Chandra et al. }            
   \titlerunning{Study of outbursts in XTE J1946+274 with \textit{AstroSat} }  

   \maketitle

%
%
\section{Introduction}           
\label{sect:intro}

Be/X-ray binary (Be/XRB) systems are transient X-ray astrophysical laboratories consisting of a neutron star, and an early Oe/Be type massive companion star (\citealt{reig2011x}). These accretion powered compact objects are known to show transient X-ray outbursts which are categorized into two classes depending on their luminosity. The X-ray outbursts in these systems are powered by the accreted matter from the decretion disc in the equatorial plane of the massive star which is formed from the ejecta from the rapidly spinning massive star (\citealt{porter2003}). The more frequent Type I outbursts are less luminous (typical luminosity $< 10^{37}$ \,erg\,s$^{-1}$) and are usually phase locked with the periastron passage of the compact object when it disrupts the disc of the Be star leading to beginning of an X-ray outburst (\citealt{cheng2014spin}). In case the decretion disc around the Be star is not completely destroyed during the first encounter with the compact object, X-ray outburst(s) occur during successive periastron passage(s) of the compact object. The Type I outbursts usually last for about 20-30 per cent of the orbital period of the binary system and are modulated at the orbital period of the system (\citealt{ziolkowski2002x}). On the other hand, Type II outbursts are more luminous (typical luminosity $> 10^{37}$ \,erg\,s$^{-1}$), rare and these outbursts are likely  triggered by episodes of sudden mass ejection from the Be star (\citealt{kriss19831980,okazaki2001natural}) but the underlying mechanism causing the sudden mass loss has remained elusive (\citealt{reig2011x,cheng2014spin}). Type II outbursts usually last for a few weeks to a few months ($\gtrsim \rm{0.5 P_{orb}}$) but sometimes they last for a few orbital periods of the system and are not locked to any particular orbital phase (\citealt{ziolkowski2002x,cheng2014spin})

XTE J1946+274 was discovered by the All Sky Monitor (ASM) onboard the \textit{Rossi X-Ray Timing Explorer} (\textit{RXTE}) (\citealt{smith1998xte}) mission. The object was localized within a 6$\degree \times$26$\degree$ error box within the ASM field of view. A new object was also detected around the same location (localized to a 5$\degree \times$8$\degree$ error box) at the same time by the Burst And Transient Source Experiment (BATSE) onboard the \textit{Compton Gamma-Ray Observatory} (\textit{CGRO}) and named GRO J1944+26 (\citealt{wilson1998xte}). Pulsations with a period of 15.83 s were detected in GRO J1944+26 by \cite{wilson1998xte} using BATSE data. Follow-up pointed observations of XTE J1946+274 with the Proportional Counter Array (PCA) aboard the \textit{RXTE} also detected 15.8 s pulsations, conﬁrming that BATSE and \textit{RXTE} had detected the same object XTE J1946+274/GRO J1944+26 (\citealt{smith1998xte}). \cite{heindl2001discovery} detected a cyclotron line at $\sim$35 keV in the energy spectrum extracted from the PCA and the High Energy X-Ray Timing Experiment (HEXTE) onboard the \textit{RXTE}. \cite{campana1999evidence} studied the \textit{RXTE}/ASM 2-10 keV monitoring observations spanning about a year since beginning of the 1998 outburst and observed a total of ﬁve outbursts. They also found hint of an 80 d modulation in the X-ray ﬂux and suggested 80 d (or double of this) to be the orbital period of the system. \cite{campana1999evidence} suggested that the first outburst in 1998 September was likely a giant (Type II) outburst as it had different rise and decay time-scales compared to the following four outbursts and was comparatively brighter by about a factor of 2. The 80 d modulation in X-ray outburst flux was suggested to be half of the $\sim$170 d orbital period (\citealt{wilson2003xte}). The binary XTE J1946+274 was also observed with the \textit{Indian X-Ray Astronomy Experiment} (\textit{IXAE}) during 1999 September 18-30 and 2000 June 28-July 7 (\citealt{paul2001ixae}). They detected 15.8 s pulsations and deduced pulse profiles in 2-6 keV and 6-18 keV bands which were found to be similar double-peaked proﬁles in both the observations. They also detected intrinsic spin-up of the pulsar during the outburst and suggested that the binary system had an eccentric orbit. The source XTE J1946+274 was in an active state during 1998 September-2001 July, after which it became dormant dropping below the detection limit of \textit{RXTE}/PCA.

The companion star of XTE J1946+274, which is optically faint (B$\sim$18.6) but bright in the infrared (H$\sim$12.1), was identified by \cite{verrecchia2002identification} and suggested to be a Be star. The distance to the source was inferred to be about 8-10 kpc based on the observed extinction (\citealt{verrecchia2002identification}). \cite{wilson2003xte} estimated the distance to the source to be d = 9.5$\pm$2.9 kpc based on the correlation between the neutron star spin-up rate and the observed ﬂux. Recent estimates using the \textit{Gaia} telescope pin down the distance to $\sim$10 kpc (\citealt{arnason2021distances}). The eccentricity and orbital inclination of the binary
system were estimated to be about 0.2-0.3 (\citealt{wilson2003xte,marcu2015transient}) and 46\degree (\citealt{marcu2015transient}) respectively. The projected semi-major axis of the system was deduced to be about 471 lt-s (\citealt{marcu2015transient}). The mass of the companion star is not constrained but is expected to be about 10-16 $M_\odot$ (\citealt{wilson2003xte}).

In 2010 a giant outburst from XTE J1946+274 was detected by the \textit{Swift}/BAT hard X-ray transient monitor (\citealt{krimm2010swift}). The hallmarks of the 2010 outburst were similar to those observed in the 1998 outburst viz. a giant outburst followed by several low intensity periodic outbursts which were not tied to any particular orbital phase. The 2010 extended outburst lasted for about a year before the source returned to quiescence in 2011 June.

 XTE J1946+274 was detected during quiescent state using the \textit{Chandra}-ACIS observations on 2013 March 12 (\citealt{arabaci2015detection}).
 \cite{arabaci2015detection} detected optical signatures of an ongoing mass ejection event from the companion star and presence of a large Be circumstellar disc which intriguingly did not fuel X-ray outbursts. \cite{tsygankov2017x} analysed the same \textit{Chandra}-ACIS observations from 2013 and suggested that the hard spectrum of XTE J1946+274 below the propeller line was most likely due to accretion from the cold disc. A single outburst, lasting for about four weeks was detected from XTE J1946+274, around 2018 June which was unlike the extended outbursts seen in 1998 and 2010. The pulsar was dormant until around 2021 September 20 (MJD 59477) when it was detected by the \textit{MAXI} mission (\citealt{nakajima2021maxi}). The 2021 outburst lasted for about four weeks after which the pulsar returned back to quiescence.

The broad-band energy spectrum of XTE J1946+274 is typical of X-ray pulsars (\citealt{filippova2005hard}). The continuum spectrum can be described by a power law with a high-energy exponential cutoff model. A cyclotron absorption line at $\sim$35 keV was detected by \cite{heindl2001discovery} using \textit{RXTE} observations during the 1998 outburst of the pulsar. The presence of a cyclotron line with energy shifting between 35 keV to 40 keV was inferred independently by \cite{doroshenko2017bepposax} using \textit{BeppoSAX} observations of the 1998 outburst. However, \cite{muller2012reawakening} ruled out the presence of a cyclotron line in 35-38 keV range using the \textit{RXTE} observations during its 2010 outburst. \cite{muller2012reawakening} instead suggested presence of a cyclotron line around 25 keV which led the authors to surmise that the cyclotron energy could possibly vary during different outbursts of the source. However, using \textit{Suzaku} data during the same outburst, \cite{maitra2013pulse} and \cite{marcu2015transient} detected presence of cyclotron absorption line around
35-38 keV but failed to detect any absorption feature around 25 keV. In a recent study, \cite{gorban2021study} detected a cyclotron absorption feature at $\sim$38 keV using \textit{NuSTAR} observations of the 2018 outburst of the pulsar.

In this paper, we investigate the timing and spectral characteristics of XTE J1946+274 using X-ray observations from the \textit{AstroSat} mission during the 2018 and 2021 outburst of this pulsar. We have used simultaneous X-ray observations from the Soft X-ray Telescope (SXT) and the Large Area X-ray Proportional Counter (LAXPC) instruments onboard \textit{AstroSat} in this study. We describe observations from the \textit{AstroSat} satellite followed by SXT and LAXPC data analysis procedures in section 2. We show results related to the timing studies of this pulsar in section 3. We derive broad-band (0.5-80 keV) energy resolved X-ray pulse profiles which is followed by results from the broad-band  spectral analysis of the pulsar. The X-ray luminosity of the source is inferred based on the broad-band spectral analysis. We explore salient features of an assemblage of outbursts shown by this source since its discovery in 1998 and discuss possible mechanisms which can drive these mysterious outbursts in this Be/XRB. We summarize our findings in section 4.

\section{Observations and data reduction}

\begin{figure}
        \centering
                \includegraphics[width=0.8\linewidth,keepaspectratio=true]{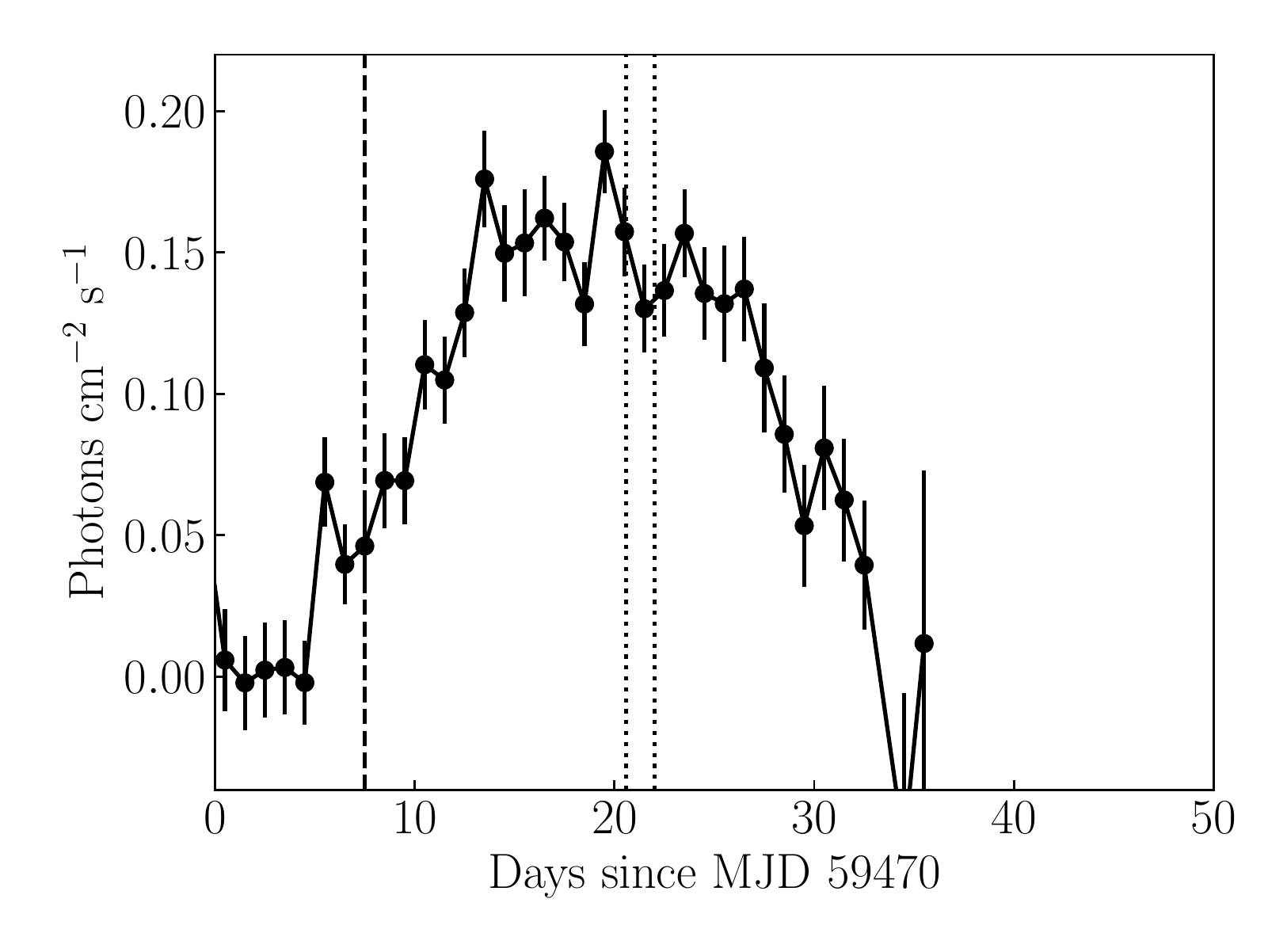}
        \caption{\textit{MAXI} one day averaged light curve of the 2021 outburst of XTE J1946+274 in the 2-20 keV energy band. The duration of overlapping \textit{AstroSat} LAXPC20 observations are shown by dotted vertical lines. The beginning of the outburst detected by the \textit{MAXI} mission is shown by a dashed vertical line.}
        \label{fig1new}
\end{figure}

Target of Opportunity (ToO) observations of XTE J1946+274 were performed by \textit{AstroSat} approximately two weeks after the source underwent a new outburst on 2021 September 20 detected by the \textit{MAXI} mission (\citealt{nakajima2021maxi}). The X-ray observations were spread over almost 39 contiguous orbits from 2021 October 3 until 2021 October 6. In this study, we have analysed data from the SXT and the LAXPC instruments covering orbits 32526-32573 yielding a total exposure of about 285 ks. The log of \textit{AstroSat} 2021 observations used in our study is shown in Table ${\ref{t1}}$. The total exposure time shown in Table ${\ref{t1}}$ for 2021 observations is about 340 ks. The recorded
data in each orbit has an overlap with the orbits before and after that orbit which is filtered
appropriately during the analysis using the LAXPC analysis software. The actual observation
time for the 2021 observations is about 285 ks as mentioned earlier. The epochs of \textit{AstroSat} observations overlapping on the \textit{MAXI} light curve during the 2021 outburst is shown in Fig. \ref{fig1new}. The reported epoch of the beginning of the 2021 outburst (\citealt{nakajima2021maxi}) is marked by a dashed vertical line in Fig. \ref{fig1new}. We have also analysed archival \textit{AstroSat} observations from the 2018 outburst of this pulsar spread over almost 18 contiguous orbits from 2018 June 9 until 2018 June 10. The log of \textit{AstroSat} 2018 observations used in our study is shown in Table ${\ref{t2}}$.

\begin{table}
\caption{Log of \textit{AstroSat} LAXPC20 observations from the 2021 outburst used in this study.} 
\label{t1}
\centering 
\begin{tabular}{c c c c c} 
\hline\hline 
S. no. & Orbit & MJD (start) & Exposure (s)  \\  
       &     &            &   \\[0.5ex]

\hline 

1	&	32526	&	59490.6	&	1146	\\
2	&	32527	&	59490.67	&	7191	\\
3	&	32528	&	59490.75	&	7148	\\
4	&	32529	&	59490.82	&	7071	\\
5	&	32530	&	59490.89	&	7202	\\
6	&	32531	&	59490.96	&	7040	\\
7	&	32534	&	59491.15	&	16822	\\
8	&	32537	&	59491.29	&	6443	\\
9	&	32538	&	59491.36	&	6889	\\
10	&	32539	&	59491.43	&	3558	\\
11	&	32540	&	59491.5	&	3558	\\
12	&	32541	&	59491.62	&	7347	\\
13	&	32542	&	59491.69	&	7270	\\
14	&	32543	&	59491.76	&	7015	\\
15	&	32544	&	59491.83	&	7087	\\
16	&	32545	&	59491.91	&	7175	\\
17	&	32546	&	59491.98	&	6919	\\
18	&	32548	&	59492.09	&	10544	\\
19	&	32549	&	59492.16	&	6587	\\
20	&	32550	&	59492.24	&	6962	\\
21	&	32552	&	59492.34	&	10313	\\
22	&	32553	&	59492.38	&	3474	\\
23	&	32554	&	59492.49	&	9392	\\
24	&	32555	&	59492.56	&	10218	\\
25	&	32556	&	59492.63	&	7294	\\
26	&	32557	&	59492.7	&	7169	\\
27	&	32558	&	59492.78	&	7106	\\
28	&	32559	&	59492.85	&	7258	\\
29	&	32560	&	59492.92	&	7222	\\
30	&	32563	&	59493.11	&	18858	\\
31	&	32564	&	59493.18	&	2309	\\
32	&	32566	&	59493.25	&	6436	\\
33	&	32567	&	59493.36	&	9793	\\
34	&	32568	&	59493.43	&	6398	\\
35	&	32569	&	59493.5	&	6831	\\
36	&	32570	&	59493.58	&	57512	\\
37	&	32571	&	59493.65	&	7239	\\
38	&	32572	&	59493.72	&	7171	\\
39	&	32573	&	59493.79	&	7073	\\

\hline 
\end{tabular}
\label{table:nonlin} 
\end{table}

\begin{table}
\caption{Log of archival \textit{AstroSat} LAXPC20 observations from the 2018 outburst used in this study.} 
\label{t2}
\centering 
\begin{tabular}{c c c c c} 
\hline\hline 
S. no. & Orbit & MJD (start) & Exposure (s)  \\  
       &     &            &   \\[0.5ex]

\hline 

1	&	14584	&	58278.49	&	272	\\
2	&	14585	&	58278.57	&	2657	\\
3	&	14586	&	58278.67	&	6016	\\
4	&	14588	&	58278.74	&	6361	\\
5	&	14589	&	58278.82	&	6713	\\
6	&	14590	&	58278.89	&	7186	\\
7	&	14591	&	58278.96	&	7123	\\
8	&	14592	&	58279.04	&	7186	\\
9	&	14593	&	58279.11	&	7209	\\
10	&	14594	&	58279.18	&	7244	\\
11	&	14595	&	58279.25	&	7185	\\
12	&	14598	&	58279.44	&	16647	\\
13	&	14599	&	58279.51	&	2345	\\
14	&	14600	&	58279.58	&	28452	\\
15	&	14602	&	58279.69	&	6062	\\
16	&	14603	&	58279.73	&	3561	\\
17	&	14604	&	58279.79	&	3561	\\
18	&	14605	&	58279.90 &	7058	\\

\hline 
\end{tabular}
\label{table:nonlin} 
\end{table}

\subsection{Soft X-ray Telescope}
The science instruments on \textit{AstroSat} and their salient features are presented in \cite{agrawal2006broad}. SXT instrument is a soft X-ray reflecting telescope sensitive in the 0.3-8 keV range onboard \textit{AstroSat} with an
effective area of $\sim$90 cm$^{2}$ at 1.5 keV. A detailed description of SXT can be found in \cite{singh2016orbit}, \cite{singh2017soft}. The 2021 SXT observations of XTE J1946+274 were in the Fast window (FW) mode while the 2018 SXT observations were done in the Photon Counting (PC) mode having coarser timing resolution of 2.3775 s. The FW mode has a time resolution of $\sim$0.3 s, free from pile-up effect and especially meant to observe bright sources. The SXT level 1 data from 39 orbits were processed using \texttt{SXTPIPELINE} version AS1SXTLevel2-1.4b\footnote{\url{http://www.tifr.res.in/~astrosat\_sxt/sxtpipeline.html}} released on 2019 January 3, to generate level2 data for each orbit. The Level 2 SXT
\newpage
data of individual orbits are merged using Julia code \url{http://astrosat-ssc.iucaa.in/sxtData}\footnote{\url{http://www.tifr.res.in/~astrosat\_sxt/dataanalysis.html}}. A circular region with 4 arcmin and 15 arcmin radius centered on the source, was used to extract source light curves and spectra for the 2021 FW mode and 2018 PC mode observations respectively. Similarly, circular regions having radius of 4 arcmin and 15 arcmin, were used to extract background light curves for the 2021 and 2018 observations respectively. Light curves and spectrum were generated using XSELECT utility in \texttt{HEASOFT} package\footnote{\url{http://heasarc.gsfc.nasa.gov/}}(version 6.30). We have used SkyBkg\_comb\_EL3p5\_Cl\_Rd16p0\_v01.pha background file and sxt\_pc\_mat\_g0to12.rmf response file provided by the SXT instrument team. Further we have used SXT arf generation tool (sxtARFModule \footnote{\url{https://www.tifr.res.in/~astrosat_sxt/dataanalysis.html}}) to generate vignetting corrected arf  ARFTESTS1\_Rad4p0\_VigCorr.arf using the arf provided for FW mode SXT arf sxt\_fw\_excl00\_v04\_20190608.arf by the SXT instrument team.

\subsection{Large Area X-ray Proportional Counter}

Large Area X-ray Proportional Counter (LAXPC) instrument onboard \textit{AstroSat} satellite, consists of 3 identical collimated detectors (LAXPC10, LAXPC20, and LAXPC30).
The arrival times of each detected X-ray photon is recorded with a time resolution of 10 $\mu$s.
The details of the characteristics of the LAXPC instrument are available in \cite{roy2016performance}, \cite{yadav2016}, and \cite{agrawal2017large}. The calibration details of LAXPC instrument are given in \cite{antia2017calibration}.
The latest calibration details of LAXPC instrument are given in \cite{antia2022astrosat}.
We have used \texttt{LAXPCSOFT}\footnote{\url{http://astrosat-ssc.iucaa.in/laxpcData}} software
to reduce Level-1 raw data file to Level-2 data. The LAXPC Level-2 data products are discussed in \cite{chandra2020study,chandra2021detection}.
The standard routines \footnote{\url{http://astrosat-ssc.iucaa.in/uploads/threadsPageNew_LAXPC.html}} available in \texttt{LAXPCSOFT} were used to generate the light curves and the energy spectrum. The LAXPC30 detector suffered abnormal gain changes and was switched off on 2018 March 8. In the third observation (O3), the LAXPC10 detector was operating at
low gain and so we have used data only from LAXPC20 detector in our study.

 We correct the X-ray photons arrival times to the solar system barycentre using the \textit{AstroSat} barycentric correction utility \textquoteleft as1bary\textquoteright. The orbit files for barycentric correction are generated using \textit{AstroSat} orbit file generator\footnote{\url{http://astrosat-ssc.iucaa.in:8080/orbitgen/}}. We have used the \texttt{HEASOFT} software package (version 6.30) for our analysis.

\section{Results and discussions}
\subsection{Timing studies}
\begin{figure}
\begin{subfigure}{0.45\textwidth}
                \includegraphics[width=\linewidth]{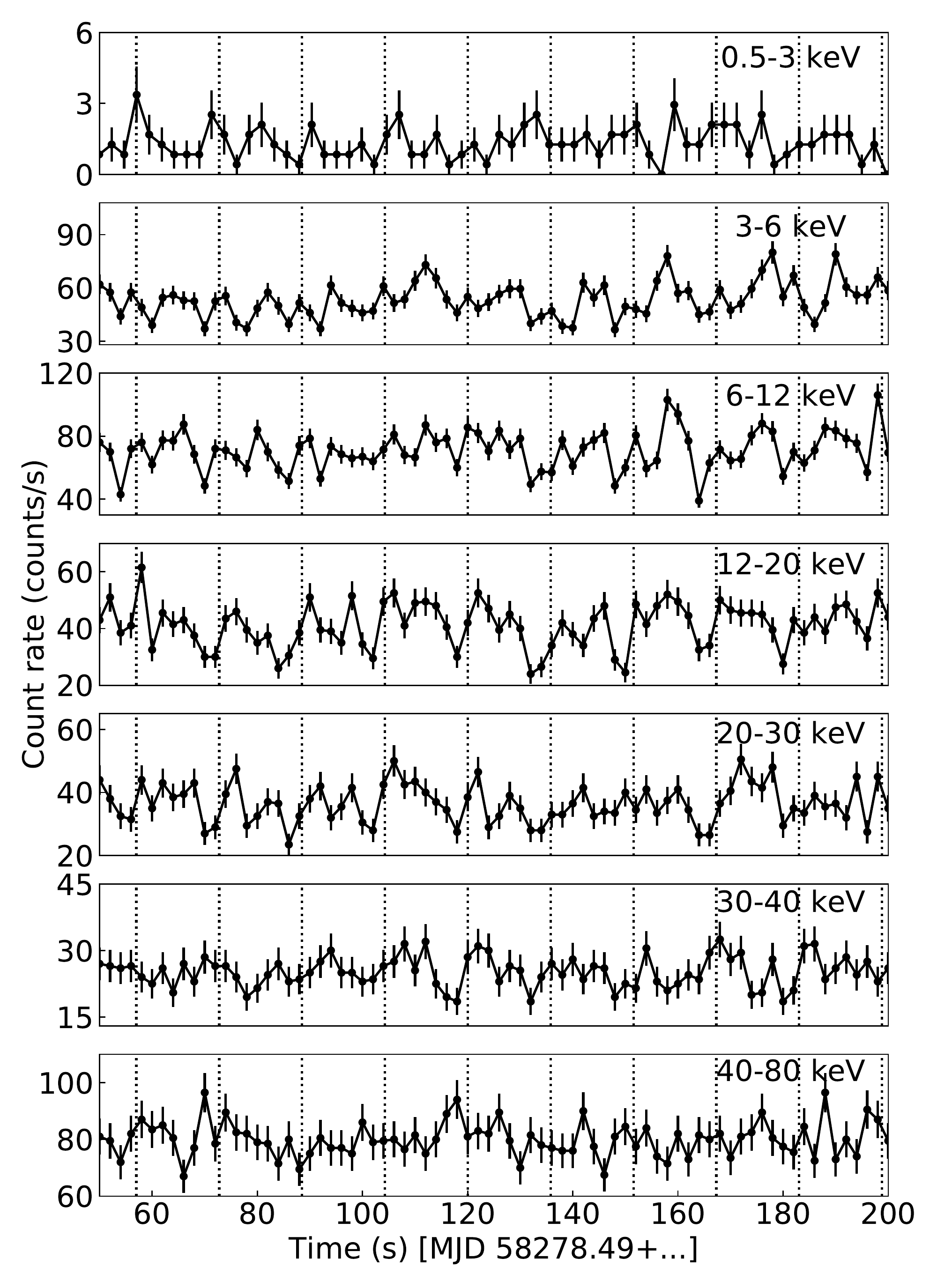}
                \caption*{(a)}
        \end{subfigure}
        \begin{subfigure}{0.45\textwidth}
                \includegraphics[width=\linewidth]{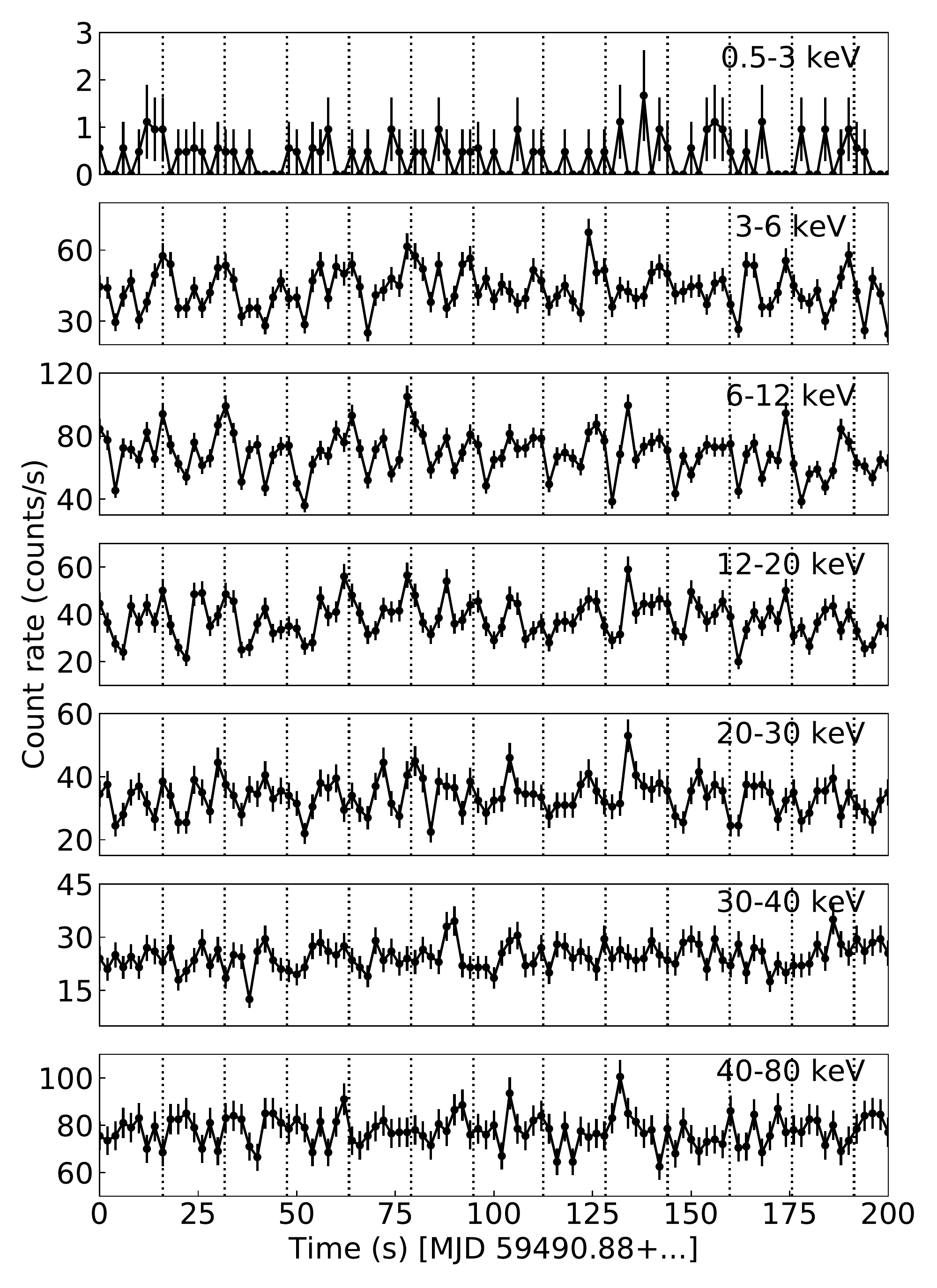}
                \caption*{(b)}
        \end{subfigure}\
        \caption{(a) Energy resolved light curves of XTE J1946+274 in the 0.5-80 keV energy band using \textit{AstroSat} SXT (0.5-3 keV band) and LAXPC20 (3-80 keV band) observations of the 2018 giant outburst. (b) Same as (a) for the 2021 giant outburst of the pulsar. All the light curves have been rebinned to 2 s. Pulsations of 15.7 s period, are visible in all the energy bands in the 0.5-80 keV energy range.}
    \label{f4}
\end{figure}

The light curves have been generated using 2 s averaged count rates in 0.5-3 keV from SXT FW mode 2021 observations while light curves in 3-6 keV, 6-12 keV, 12-20 keV, 20-30 keV, 30-40 keV, and 40-80 keV energy bands were derived from LAXPC20 observations (Fig. $\ref{f4}$). Broad-band X-ray pulsations ($P\sim 15.7 s$) are clearly seen in the light curves. Broad-band (0.5-80 keV) energy resolved pulsations are also observed in the light curves generated using the 2018 \textit{AstroSat} observations of this pulsar.

\begin{figure}
\centering
  \includegraphics[width=\linewidth]{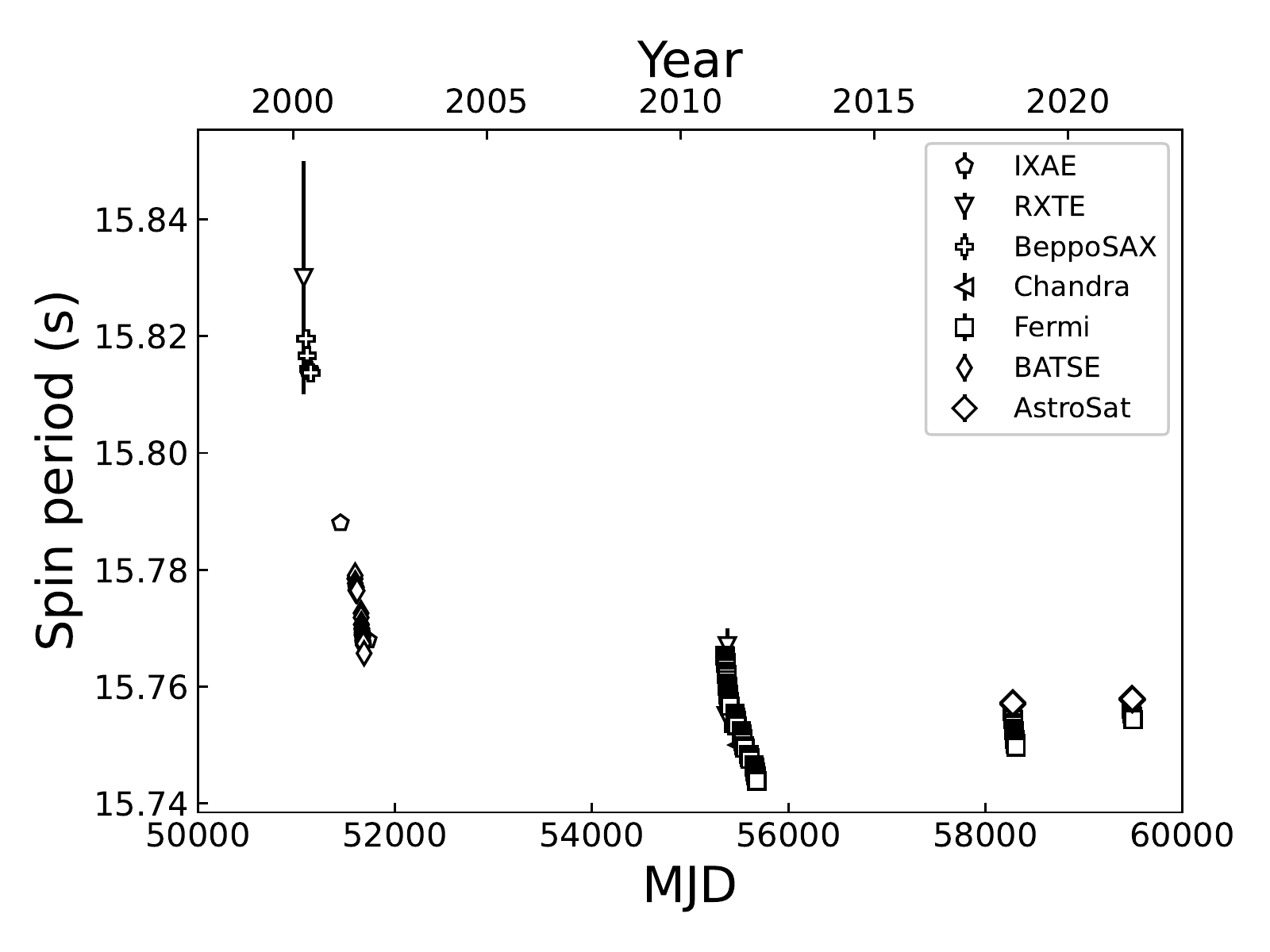}
  \caption{Long-term spin history of XTE J1946+274 from 1998 September till 2021 October. The spin-up of the pulsar during the long outbursts in 1998, 2010 and shorter outbursts in 2018 and 2021, is clearly seen. The different markers indicate spin periods inferred from  different observatories.}
 \label{f5}
\end{figure}

We use the FTOOLS subroutine \textit{efsearch} to obtain the best estimated pulse period of XTE J1946+274 from the entire 285 ks LAXPC20 2021 observations in the 3-20 keV energy band. For this timing analysis we have extracted data only from the top layer of the LAXPC20 detector to get the best signal-to-noise ratio. The pulse period obtained from our timing analysis is 15.757818$\pm$0.000040 s. The pulsation peak inferred using efsearch was fitted with a Gaussian whose width provided the error on the estimated period. Similarly, the best estimated pulse period of XTE J1946+274
from the entire 140 ks LAXPC20 2018 observations is 15.757119$\pm$0.000086 s. Thereafter, we investigate the temporal evolution of the spin period of the source by estimating the pulse period for each 15 ks successive LAXPC20 observation segments and construct the evolution of the pulsar spin period since its discovery in 1998 which is shown in Fig. $\ref{f5}$.

It is observed that the pulsar spins-up during the outbursts but switches to spin-down state in the quiescent periods between the outbursts. This is consistent with the general finding of \cite{malacaria2020ups} based on monitoring the pulsation periods of 39 accreting pulsar binaries, of which
28 are Be pulsars with the \textit{Fermi}/Gamma Ray Burst Monitor (\textit{Fermi}/GBM, \cite{meegan2009fermi}), that almost all the pulsars in the Be binaries undergo rapid spin-up during the outburst due to high accretion torque but at the end of the outburst they again switch to spin-down mode. It is also found that the spin-up rates are higher (factor of $\sim$2 to 5), compared to the spin-down rates. Over a long time the net
spin period evolution exhibits a spin-up though at a very low rate of $\sim 10^{-14}$ \,s \,s$^{-1}$. We estimate the changes in spin rates between the outbursts assuming that the pulsar spin-down rate is monotonic during the dormant periods. The estimated rate of spin changes during the dormant periods MJD 51727.5-55353, MJD 55677-58275 and MJD 58311-59481 are  $\sim-8.26 \times 10^{-12}$ \,s \,s$^{-1}$, $\sim5.27 \times 10^{-11}$ \,s \,s$^{-1}$ and $\sim6.89 \times 10^{-11}$ \,s \,s$^{-1}$ respectively. Note that during the period MJD 51727.5-55353 the pulsar is in a
dormant state, and yet the pulsar is showing a modest spin-up contrary to the general trend. However, the spin-up rate is an order of magnitude smaller compared to the spin-up rates generally deduced during the outbursts.

\begin{figure}
\centering
  \includegraphics[width=\linewidth]{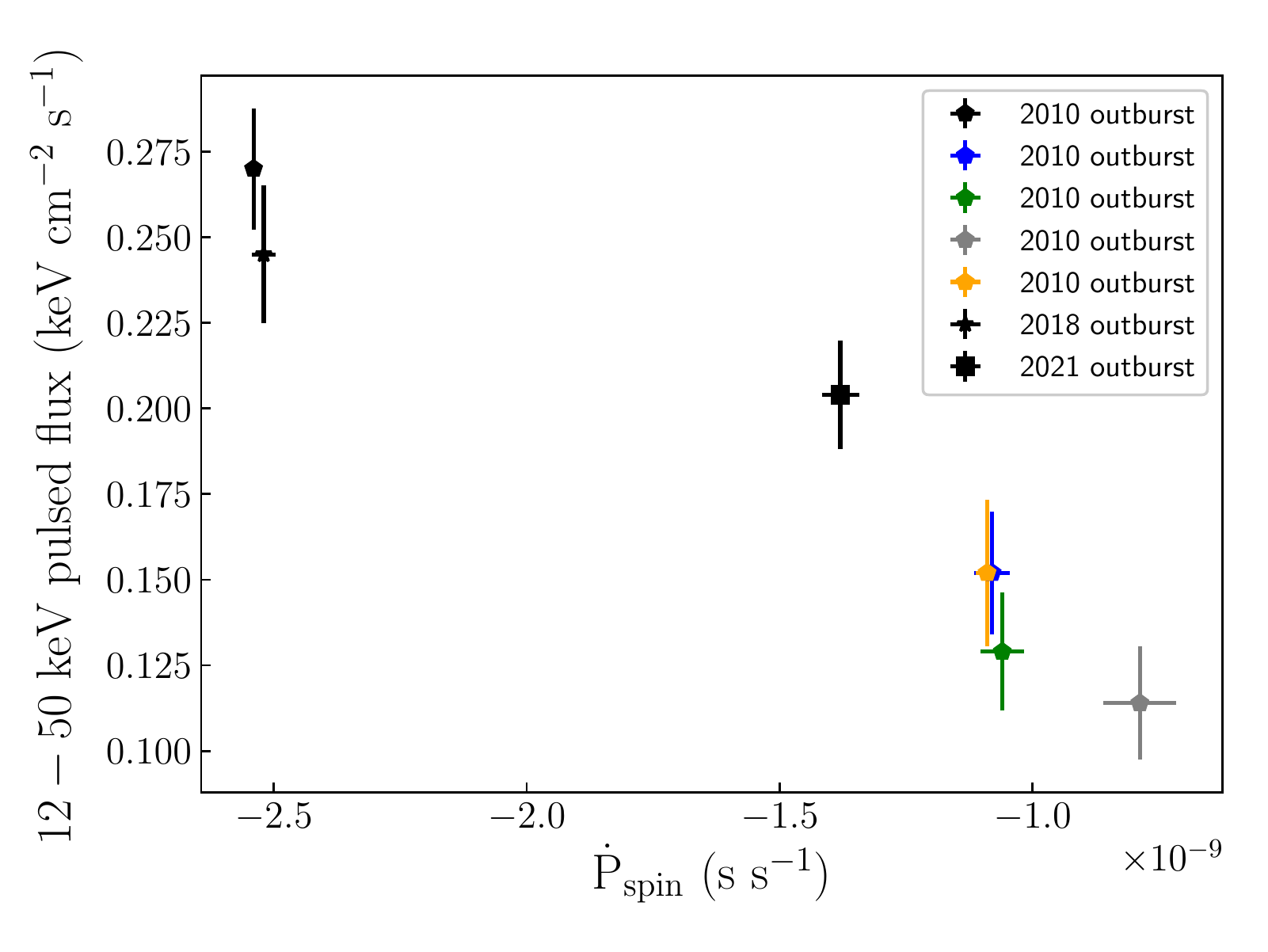}
  \caption{Plot of spin period derivative vs 12-50 keV pulsed flux from \textit{Fermi}/GBM observations of the pulsar. The best fit estimate of the spin period derivative has been derived from linear fit to the data of spin evolution of the pulsar from \textit{Fermi}/GBM observations. Values derived from 2010, 2018 and 2021 outbursts have been shown by pentagon, star and squares. The values of the two parameters measured from the series of bursts from 2010 are shown in different colours. The first, second, third, fourth and fifth bursts have been shown in black, blue, green, gray and orange colours respectively.}
 \label{f7}
\end{figure}

Fig. \ref{f7} shows comparison between the spin period derivative and the 12-50 keV pulsed flux from \textit{Fermi}/GBM observations of this pulsar. The values derived from series of outbursts from 2010 are shown in different colours. The values derived from the first, second, third, fourth and fifth outbursts have been shown in black, blue, green, gray and orange colours. There is suggestion of a linear trend between the pulsed flux in the 12-50 keV band and the spin-up rate.
Based on correlation between the spin-up rate and flux it has been suggested that an accretion disc was present during the 1998 outburst of this pulsar (\citealt{wilson2003xte}).
 It is also observed from Fig. \ref{f7} that the maximum 12-50 keV pulsed flux during the giant outburst in 2010 was about twofold more than that during the following four smaller outbursts. Interestingly, the maximum pulsed flux in the 12-50 keV energy band during the 2018 and 2021 outbursts of this pulsar are comparable to that during the giant outburst in 2010 suggesting that the underlying mechanism for these outbursts might be similar.
Interestingly, from the estimated spin-up rates using \textit{Fermi}/GBM observations, the low intensity outbursts have spin-up rate about a factor of 2 smaller than that estimated for giant outbursts.

\subsection{Broad-band energy resolved pulse profiles}

\begin{figure*}
        \begin{subfigure}{0.48\textwidth}
                \includegraphics[width=\linewidth]{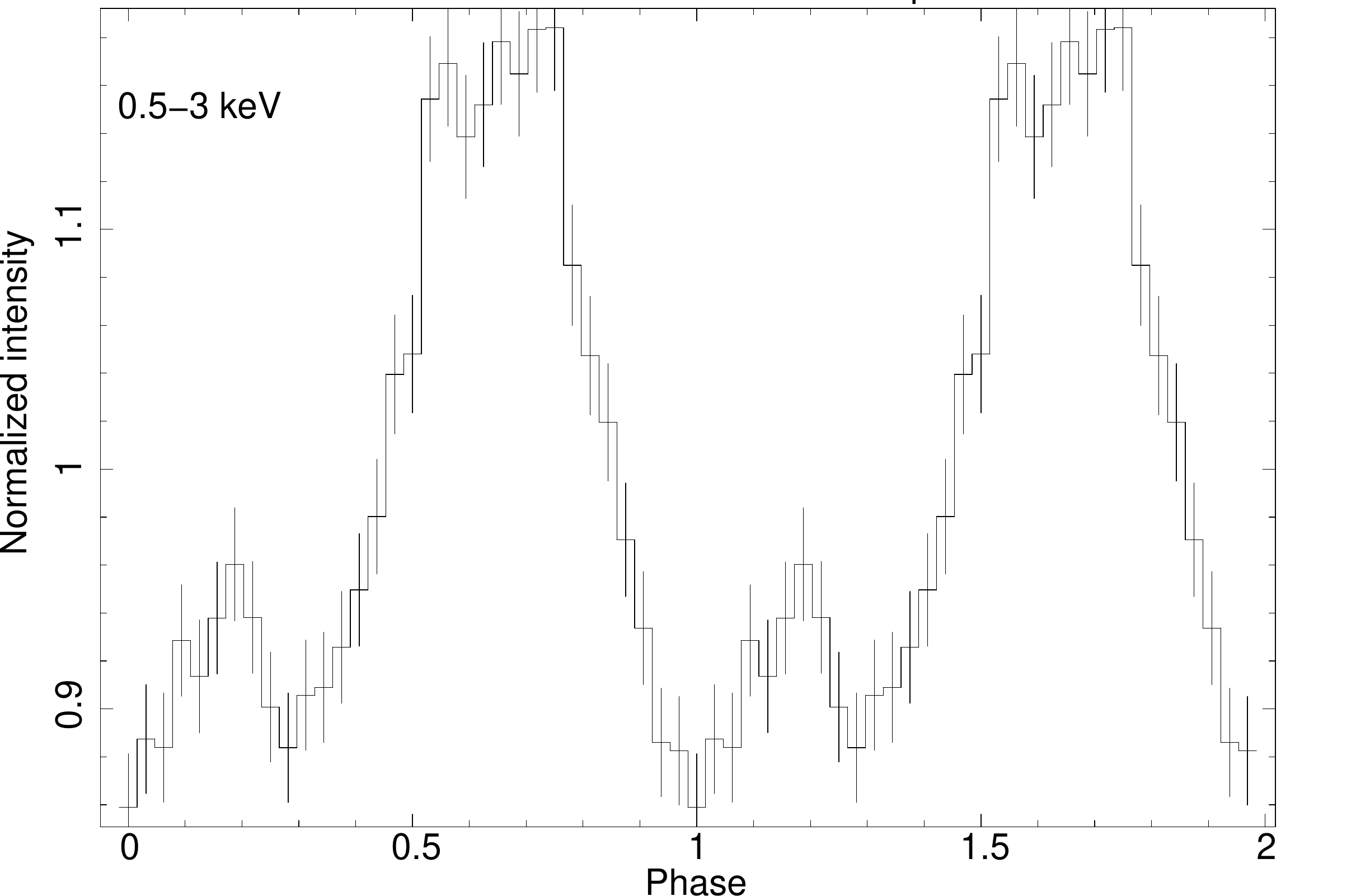}
                \caption*{(a)}
                \includegraphics[width=\linewidth]{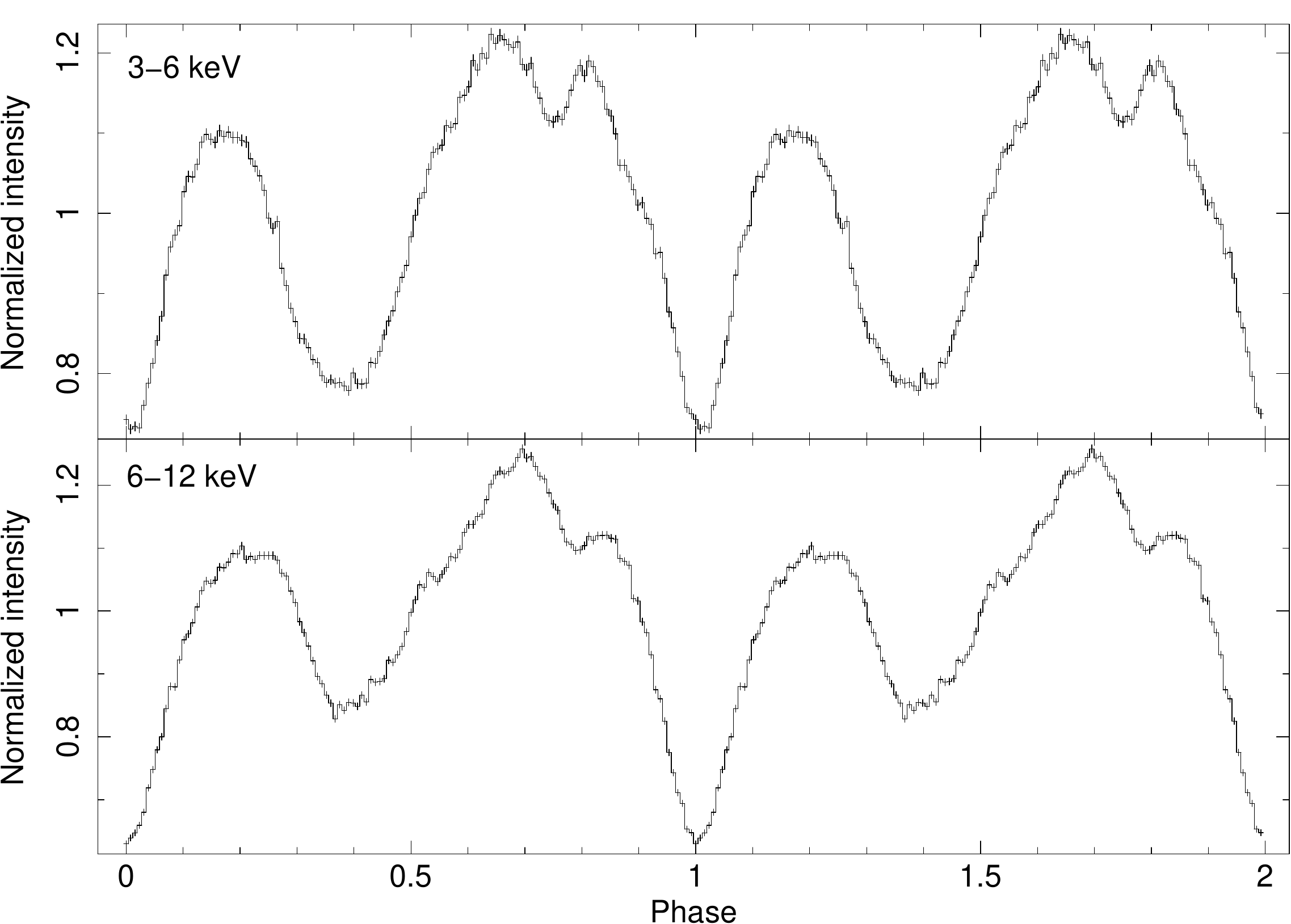}
                \caption*{(b)}
                \includegraphics[width=\linewidth]{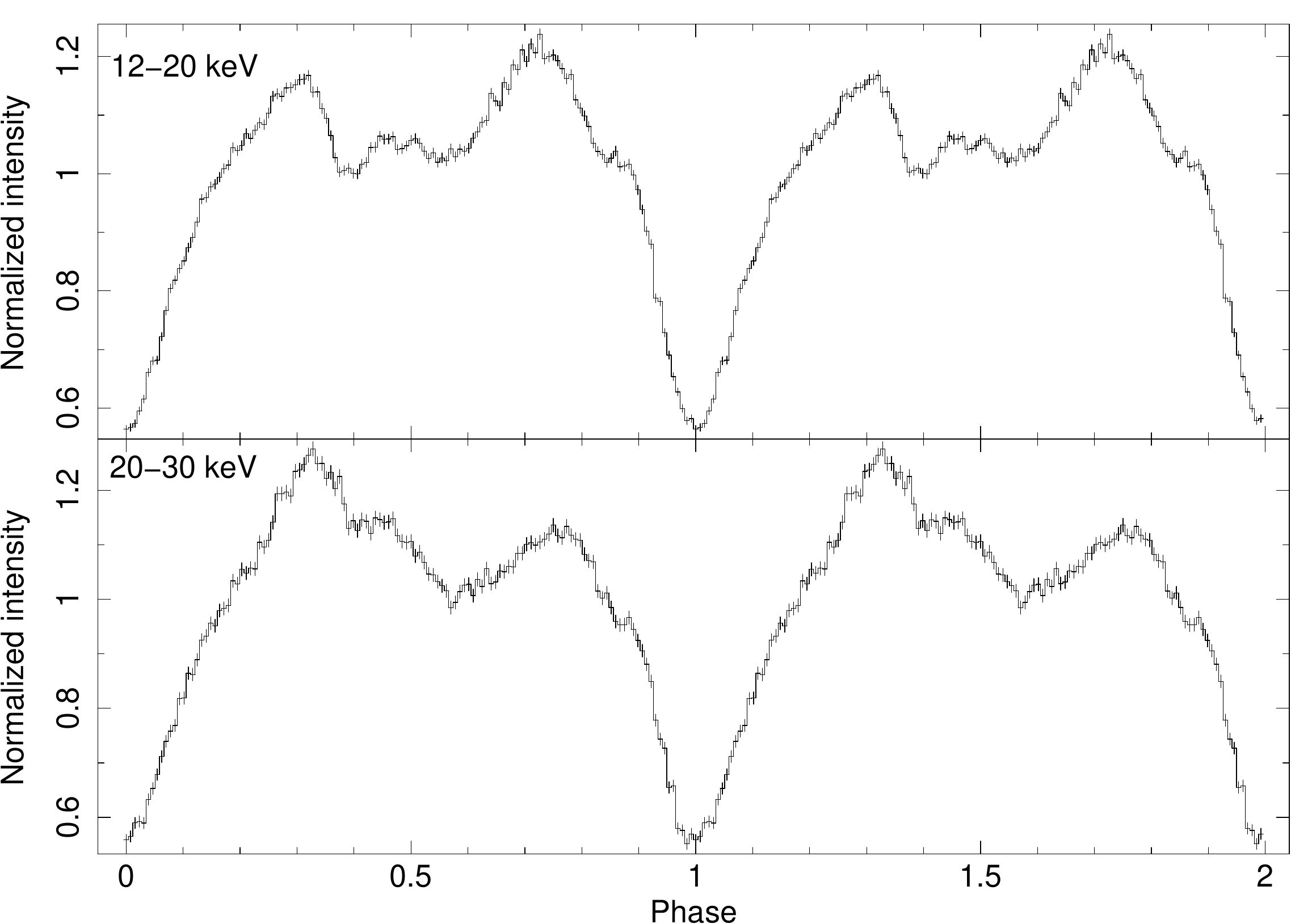}
                \caption*{(c)}
                \includegraphics[width=\linewidth]{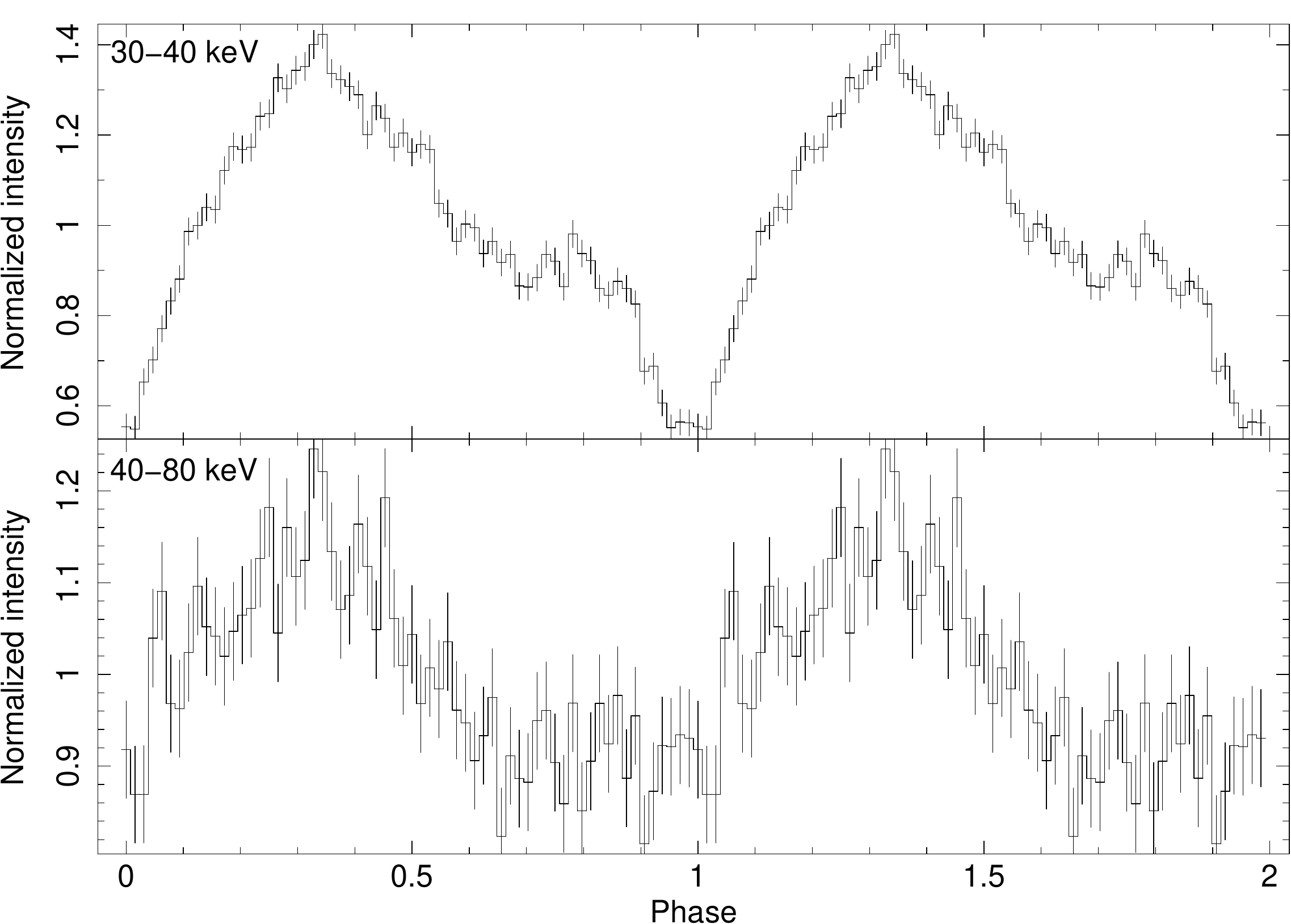}
                \caption*{(d)}
        \end{subfigure}
        \begin{subfigure}{0.48\textwidth}
                \includegraphics[width=\linewidth]{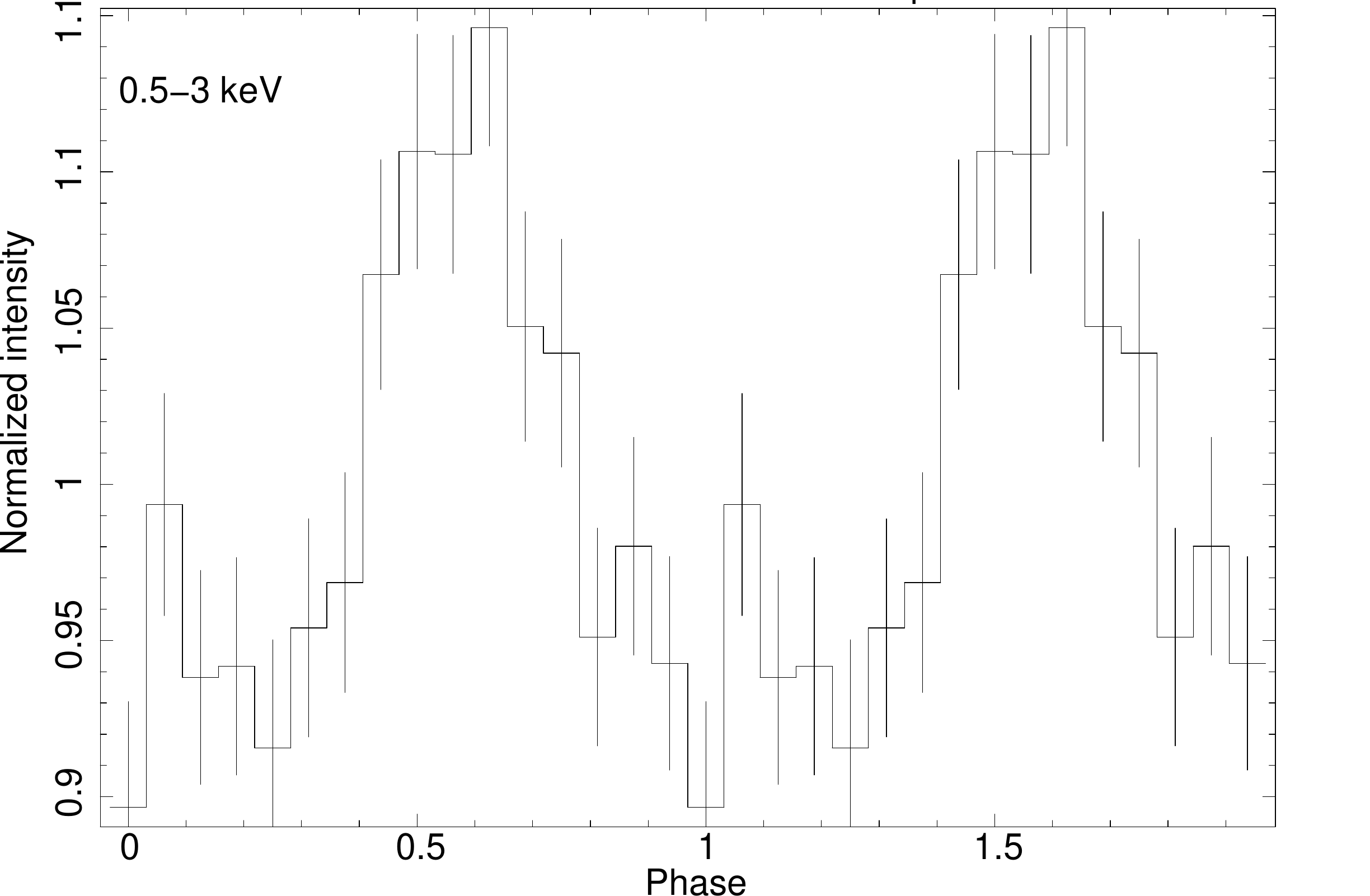}
                \caption*{(d)}
                \includegraphics[width=\linewidth]{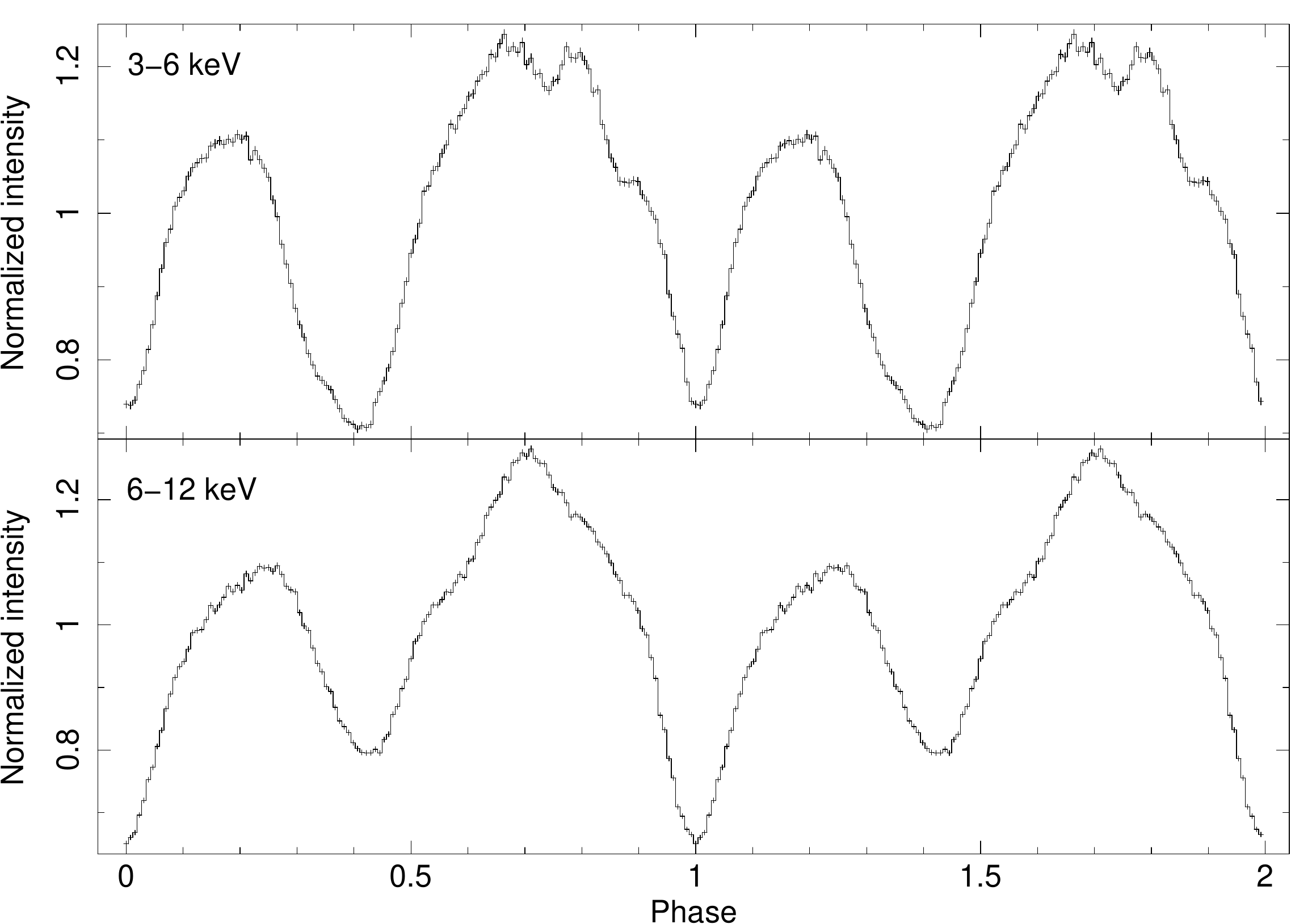}
                \caption*{(e)}
                \includegraphics[width=\linewidth]{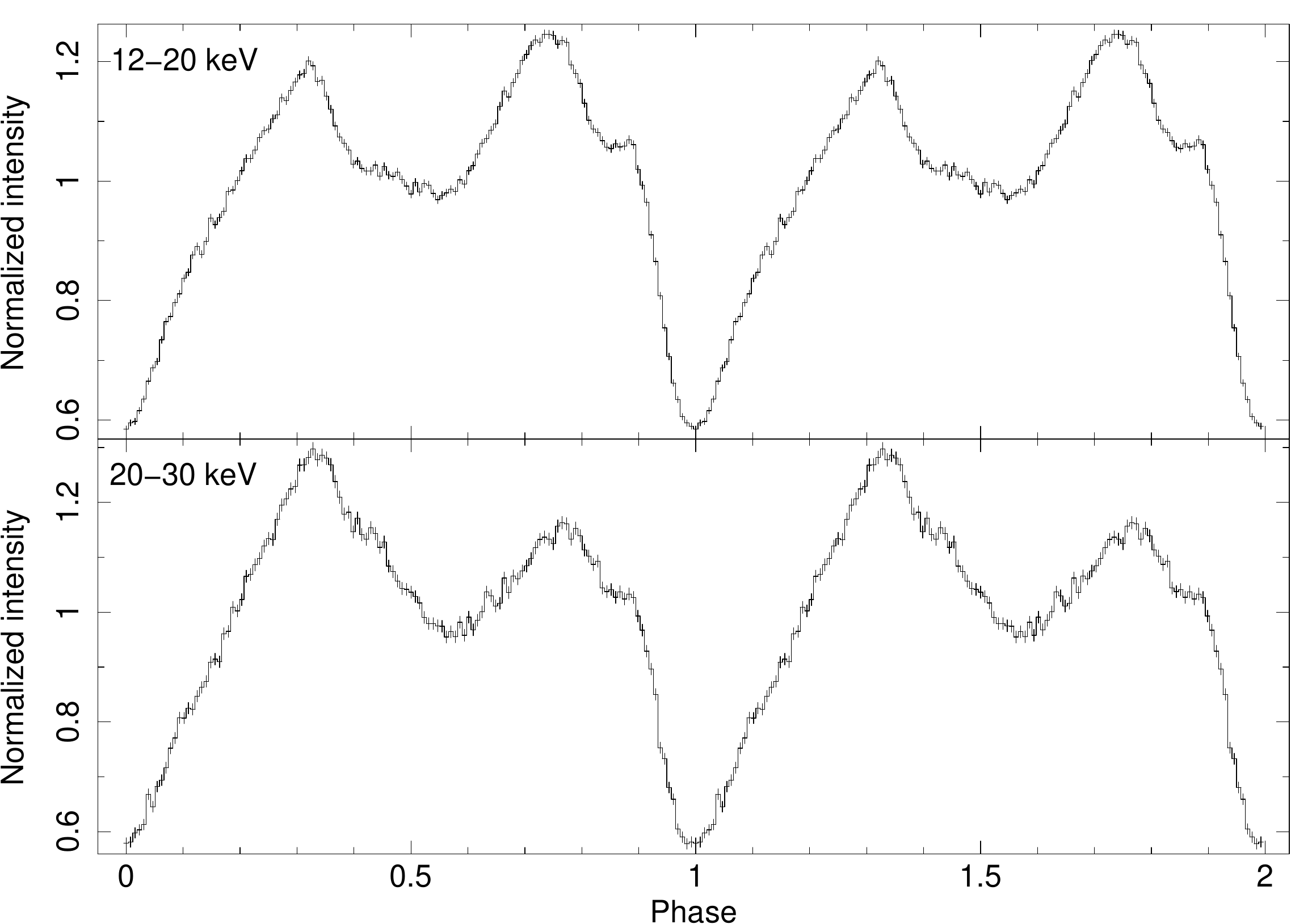}
                \caption*{(f)}
                \includegraphics[width=\linewidth]{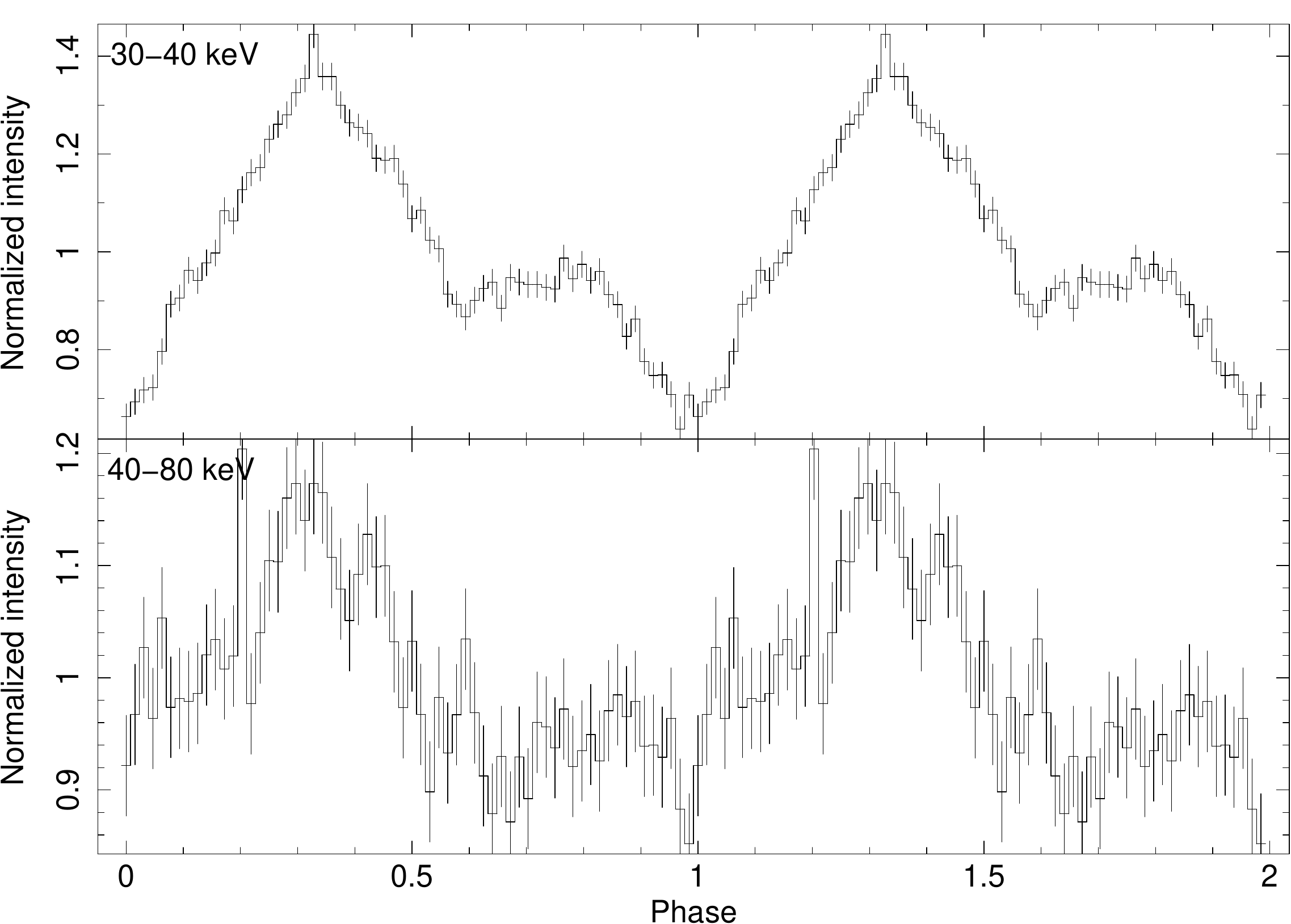}
                \caption*{(g)}
        \end{subfigure}\
        \caption{\small{Folded pulse profiles of XTE J1946+274 from the 2018 outburst are shown in different energy ranges (a) 0.5-3 keV (b) 3-6 keV and 6-12 keV (c) 12-20 and 20-30 keV (d) 30-40 keV and 40-80 keV. The folded profiles in 0.5-3 keV, 3-30 keV and 30-80 keV energy range are binned in 32 bins, 128 bins and 64 bins respectively using MJD 58278.49 as the epoch. Similar pulse profiles constructed from the 2021 outburst are shown in the same energy bands on the right side of the figure. The epoch used for folding profiles from 2021 observations is MJD 59490.57.}}
        \label{f8}
\end{figure*}

The background subtracted folded profiles in the 0.5-3 keV, 3-6 keV, 6-12 keV, 12-20 keV, 20-30 keV, 30-40 keV, and 40-80 keV energy band obtained from SXT and LAXPC20 observations from the 2018 and 2021 outbursts of this source are shown in Fig. $\ref{f8}$. We have extracted LAXPC20 data only from the top layer for 3-6 keV, 6-12 keV, and 12-20 keV energy band while data from all the layers have been extracted for 20-30 keV, 30-40 keV, and 40-80 keV energy range. Broadly, the
pulse profiles in the various energy bands appear similar for the 2018 and 2021 outbursts. It is, however, clearly observed that the pulse profiles evolve with energy in both the outbursts as also observed during earlier outbursts of this source (\citealt{wilson2003xte, maitra2013pulse, marcu2015transient,arabaci2015detection,doroshenko2017bepposax,gorban2021study}). On closer examination, one finds noticeable differences in the pulse shape and its amplitude. During the 2018 outburst, the pulse shape in 0.5-3 keV is composed of two peaks with the second peak
relatively stronger than the first peak. The first peak has an asymmetric leading side while the leading side of the second peak has comparatively less asymmetricity. During the 2021 outburst, the signal-to-noise ratio (SNR) of the profile in this energy band is comparatively less and although two distinct peaks are observed, the asymmetricity on the leading edge of the first peak seems to have disappeared.

\begin{figure*}
        \begin{subfigure}{0.45\textwidth}
                \includegraphics[width=\linewidth]{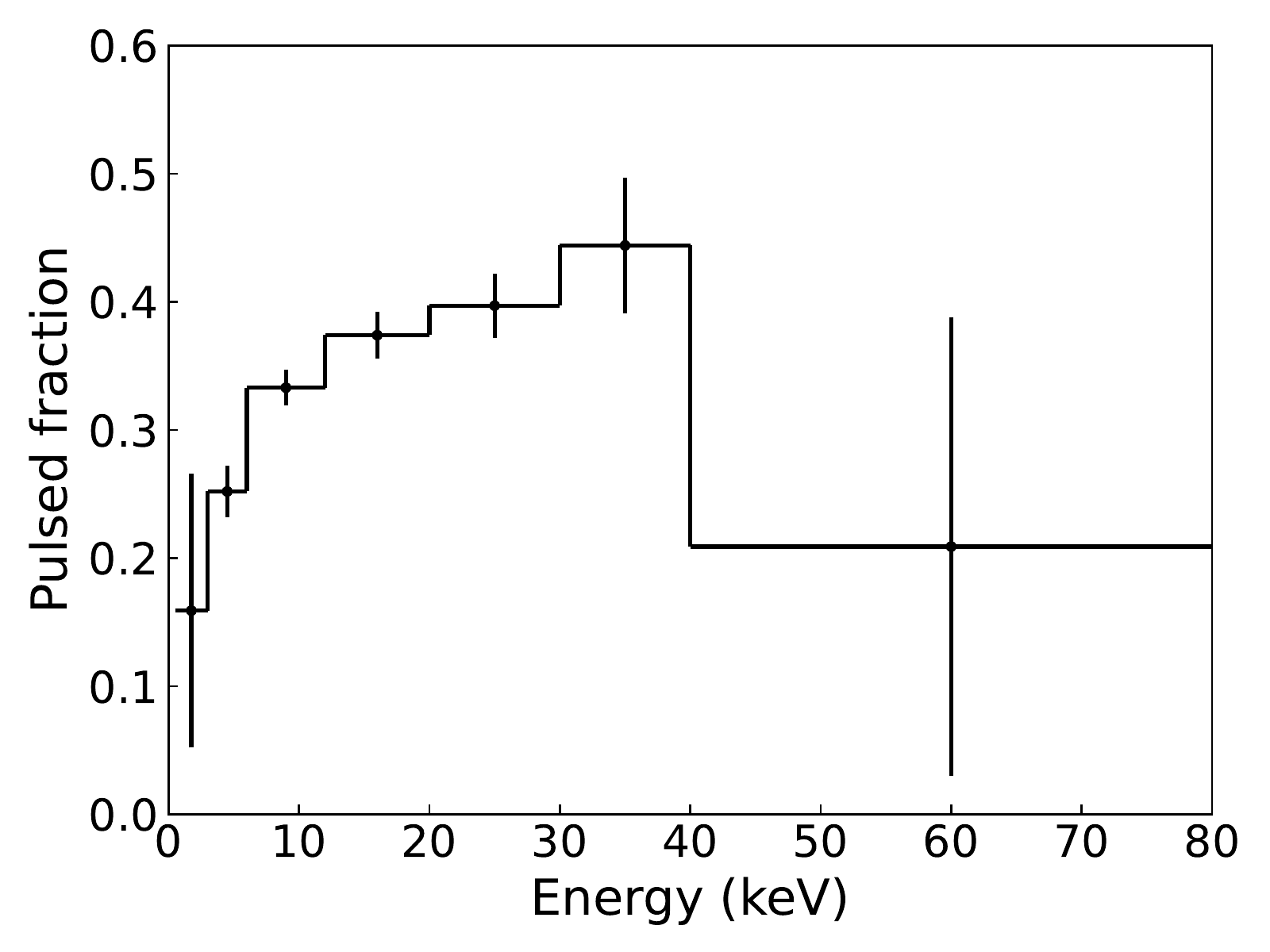}
        \end{subfigure}
        \begin{subfigure}{0.45\textwidth}
                \includegraphics[width=\linewidth]{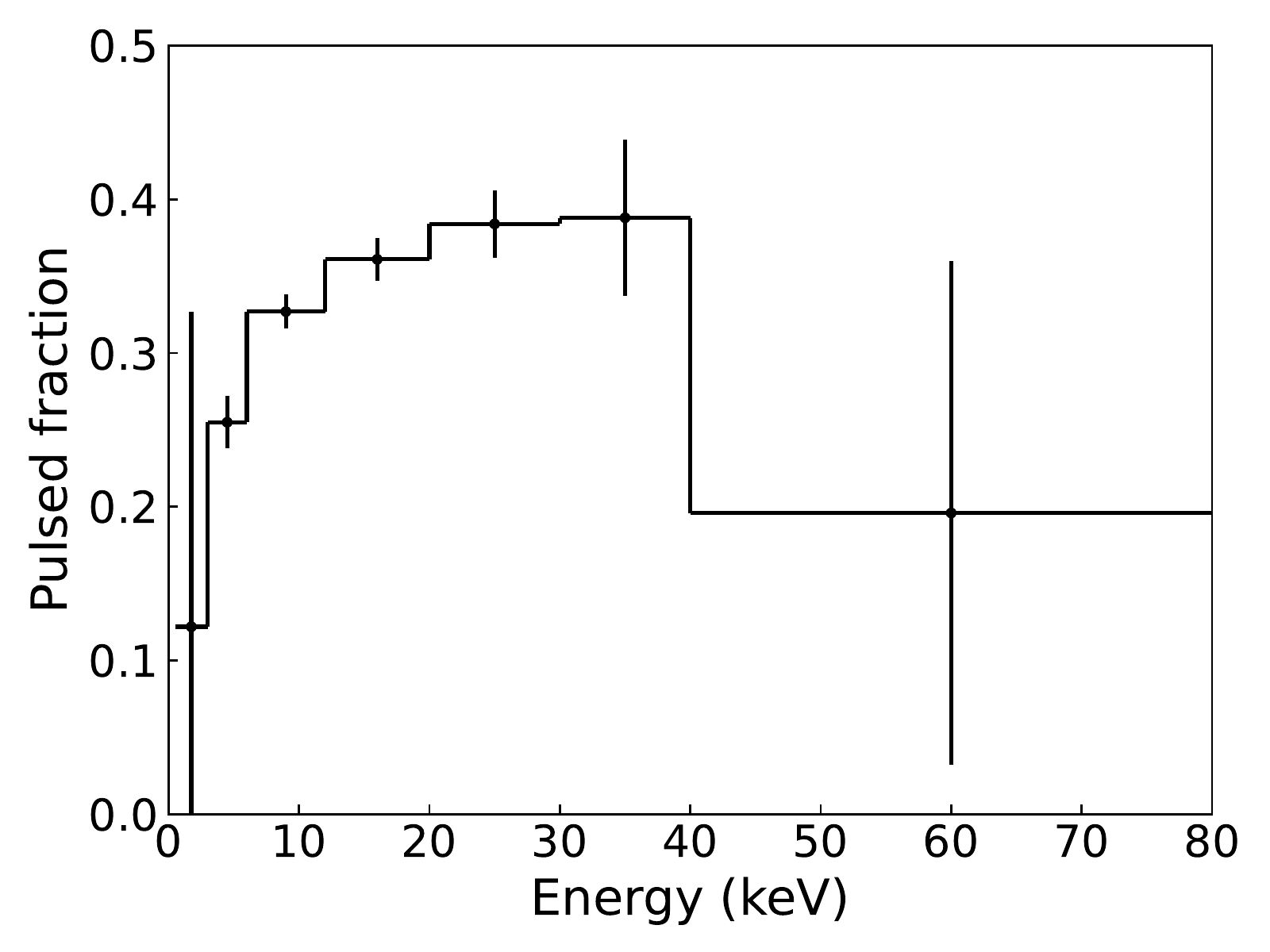}
        \end{subfigure}\
        \caption{Pulsed fraction in different energy ranges versus energy inferred from SXT and LAXPC20 observations of the 2018 and 2021 outburst of the pulsar shown in the left and the right panel respectively.}
        \label{f9}
\end{figure*}

In both the outbursts, as the energy increases, the first peak gradually becomes relatively stronger, almost becomes comparable to the amplitude of the second peak in the 12-20 keV energy band and becomes the dominant peak above 20 keV. This transition in relative amplitudes of  the two peaks has been observed during the 1998 outburst (\citealt{wilson2003xte,doroshenko2017bepposax}), 2001 outburst (\citealt{wilson2003xte}), 2010 outburst (\citealt{maitra2013pulse,marcu2015transient}) and 2018 outburst (\citealt{gorban2021study}) of the source. The separation in phase between the two peaks is around 0.45 which maybe attributed to   emission from the opposite poles of the neutron star. Interestingly, the phase separation of $\sim$0.45 is maintained as the two peaks slowly move towards later phases which is clearly observed in the 3-30 keV energy band.
An additional emission component is clearly visible as a bump on the falling edge of the profile around pulse phase of 0.9 which is unclear during the 2018 outburst.
Indications of weak emission components are discernible in the saddle between the two peaks in the 12-20 keV energy band and on the falling edge of second peak in the 6-12 keV energy band which is prominent in the 2018 outburst.
The minimum between the two peaks observed around phase $\sim$0.4 becomes less shallow with increasing energy upto 20 keV and gradually disappears above 30 keV. Intriguingly, the minimum shifts to around phase $\sim$0.6 in the 20-30 keV energy range. This shift in the minimum between the two peaks can also be observed during the 1998 outburst (\citealt{doroshenko2017bepposax}), 2001 outburst (\citealt{wilson2003xte}) and 2018 outburst (\citealt{gorban2021study}) of this source.  Intriguingly, almost no shift in the minimum with energy is observed during the peak of the 1998 outburst of the source (\citealt{wilson2003xte}).
The asymmetric nature of the first peak on the leading edge persists over a broad-band energy range from 0.5-80 keV although the asymmetricity slightly decreases in the 3-12 keV energy band.

We have investigated dependence of the Pulsed Fraction (PF) defined as $\mathrm{PF}=(\mathrm{I}_{\mathrm{max}}-\mathrm{I}_{\mathrm{min}})/(\mathrm{I}_{\mathrm{max}}+\mathrm{I}_{\mathrm{min}}$) where $\mathrm{I}_{\mathrm{max}}$ and $\mathrm{I}_{\mathrm{min}}$ are the maximum and minimum intensities in the folded profile. Plots of PF vs energy are shown in Fig. $\ref{f9}$ for the 2018 and 2021 outbursts. The PF is highly energy dependent and varies with energy as shown in Fig. $\ref{f9}$. The pulsed fraction increases with energy
during both 2018 and 2021 outbursts and peaks in the energy band of 30-40 keV before decreasing at higher energies.  However, during the 2018 outburst, the increase in pulsed fraction from 0.5-3 keV to 30-40 keV energy band is gradual while during the 2021 outburst a slight jump in pulsed fraction from 0.5-3 keV to 3-6 keV energy band is observed. Enhancement in the pulsed fraction with energy has also been observed in several X-ray pulsars and a geometrical model has been proposed to explain the observed manifestation (\citealt{lutovinov2009}). However, there is some hint of decrease in the pulsed fraction above 40 keV which maybe attributed to the morphology of the accretion column of the pulsar. The asymmetricity of folded profiles in X-ray pulsars have been ascribed to various factors such as antipodal magnetic poles (\citealt{parmar1989transient, leahy1991modelling, riffert1993fitting, bulik1995geometry,kraus1995, sasaki2012}) or multi-polar magnetic fields (\citealt{greenhill1998}) or the asymmetric accretion stream in the vicinity of the neutron star (\citealt{basko1976, wang1981, miller1996}).

\subsection{Spectral studies}
\subsubsection{Spectral studies using 2018 observations}
\begin{figure}
       \centering
      \includegraphics[width=0.8\linewidth]{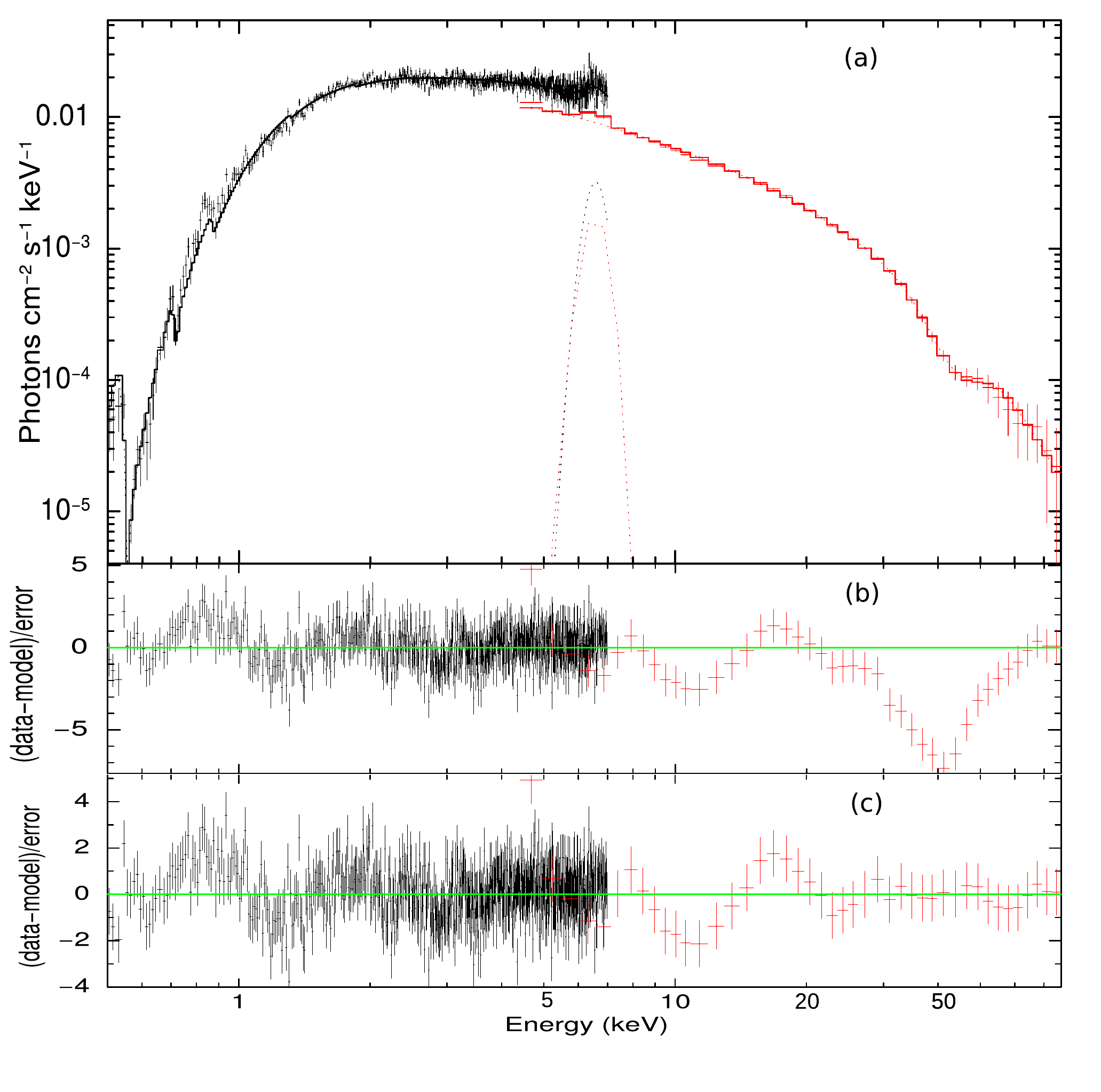}
    \caption{ (a) Simultaneous fitted SXT and LAXPC spectrum of 2018 outburst using the power law with high energy cut-off model. The best-fitting model is shown by the solid line along with the spectral data. (b) The residuals between the data and the model are shown without fitting for any cyclotron line absorption features. Prominent absorption residuals around 43 keV are clearly seen in the panel. (c) The residuals between the data and the model are shown after fitting a cyclotron line around 43 keV. The cyclotron line is detected with a significance of more than 3$\sigma$.}
    \label{f10}
\end{figure}

We have extracted the energy spectra for 2018 and 2021 outbursts by using the \texttt{LAXPCSOFT} software for the entire observation periods.
We have performed a combined spectral fitting of SXT
and LAXPC20 spectra using \texttt{XSPEC} 12.12.1 (\citealt{arnaud1996astronomical}) in the energy range 0.5-80 keV for the 2018 outburst (Fig. $\ref{f10}$). The spectral fitting of combined SXT and LAXPC data is confined to 0.5-80 keV due to reliable spectral response in this energy range. A 2 per cent systematics was included in the spectral analysis to take care of uncertainties in the response matrix. The broad-band spectrum (0.5-80 keV) is fitted using the HIGHECUT, NEWHCUT, NPEX (\citealt{mihara1995observational}) and the FDCUT (\citealt{tanaka1986observations}) models available in \texttt{XSPEC}.
We have used \textit{tbabs} model (\citealt{wilms2000}) to take care of the broad-band absorption in the spectrum during spectral fitting. The abundance and atomic cross-section used in the usage of the photoionization model (\textit{tbabs}) are angr and vern respectively. In addition, the
iron emission line at $\sim$6.6 keV was added to the combined best-ﬁt model. The results of the broad-band spectral analysis for the 2018 observations are given in table \ref{t4}. A gain correction has been applied to the SXT spectrum using the \texttt{XSPEC} command \textit{gain fit} where the slope was frozen at unity and the offset obtained from the fit are given in table \ref{t4}. A constant factor has been included in the model to allow for cross calibration difference between the SXT and LAXPC spectrum. We freeze the constant factor at unity for the LAXPC spectrum while allow this factor to
vary for the SXT spectrum. The spectrum of the pulsar during the outburst is found to be slightly hard ($\Gamma \sim 0.5$). The absorption $\mathrm{N_H}$ is deduced to be $\sim 1\times 10^{22}$ \,cm$^{-2}$. We observe deviations in the fitted spectrum around 43 keV (Fig. \ref{f10}). An absorption line with a Gaussian optical depth proﬁle \textit{gabs} was added to the model. This model accounts for cyclotron resonant scattering features (CRSFs) in the
spectrum of X-ray pulsars.

From the best fit spectrum the unabsorbed X-ray flux of XTE J1946+274 in the 0.5-80 keV energy interval in 2018 outburst is deduced to be in the range of $\sim 2.1-2.5 \times 10^{-9}$ \,erg\,cm$^{-2}$\,s$^{-1}$ which implies X-ray luminosity to be in the range of $\sim 2.3-2.7 \times 10^{37}$\,erg\,s$^{-1}$ for a distance of 9.5 kpc for the source.

\begin{table*}
\caption{\normalsize{SXT and LAXPC simultaneous spectral-fit results for XTE J1946+274 using data from 2018 \textit{AstroSat} observations of XTE J1946+274.}} 
\label{t4}
\begin{tabular}{l c r r r r}
\hline\hline 
\normalsize{Model} & \normalsize{Parameter} & \normalsize{HIGHECUT} & \normalsize{NEWHCUT} & \normalsize{NPEX} & \normalsize{FDCUT}\\
\hline 
constant & LAXPC spectrum& 1.0 (fixed)& 1.0 (fixed)& 1.0 (fixed) & 1.0 (fixed)\\

\normalsize{constant} & \normalsize{SXT spectrum} & \normalsize{$\sim 1.5$}& \normalsize{$\sim 1.6$} & \normalsize{$\sim 0.85$} & \normalsize{$\sim 1.55$}\\

\normalsize{gain offset} & \normalsize{SXT spectrum} & \normalsize{$\sim 29 ~eV$}& \normalsize{$\sim 29 ~eV$} & \normalsize{$\sim 27 ~eV$} & \normalsize{$\sim 22 ~eV$}\\


\normalsize{tbabs} & \normalsize{$\mathrm{N_{H}[10^{22}}$ \,cm$^{-2}]$} & \normalsize{$0.97^{+0.02}_{-0.02}$}& \normalsize{$0.93^{+0.05}_{-0.04}$}& \normalsize{$0.92^{+0.04}_{-0.05}$} & \normalsize{$1.03^{+0.04}_{-0.05}$}\\

\normalsize{powerlaw}    &  \normalsize{$\Gamma$} & \normalsize{$0.50^{+0.01}_{-0.01}$} & \normalsize{$0.44^{+0.03}_{-0.03}$} & \normalsize{$0.15^{+0.04}_{-0.05}$} & \normalsize{$0.49^{+0.09}_{-0.08}$}\\

\normalsize{highecut}  &   \normalsize{$\mathrm{E}_{\mathrm{cut}}$[keV]} & \normalsize{$4.92^{+0.26}_{-0.23}$} & \normalsize{4.92 (fixed)} & \normalsize{$6.45^{+0.55}_{-0.34}$} & \normalsize{$2.17^{+7.02}_{-5.00}$}\\

 & \normalsize{$\mathrm{E}_{\mathrm{fold}}$[keV]} & \normalsize{$13.81^{+0.43}_{-0.37}$} & \normalsize{$13.44^{+0.54}_{-0.47}$} & \normalsize{-} & \normalsize{$11.27^{+0.84}_{-0.96}$} \\

\normalsize{Gaussian}  & \normalsize{$\mathrm{E(Fe} ~\mathrm{line})$[keV]} & \normalsize{$6.57^{+0.12}_{-0.12}$} & \normalsize{$6.72^{+0.18}_{-0.16}$} &\normalsize{$6.33^{+0.26}_{-0.28}$} & \normalsize{$6.22^{+0.28}_{-0.27}$}\\

 & \normalsize{$\sigma(\mathrm{Fe} ~\mathrm{line})$[keV]} & \normalsize{$0.35^{+0.20}_{-0.10}$} & \normalsize{$0.5^{+0.8}_{-0.2}$} & \normalsize{$1.40^{+0.25}_{-0.22}$} & \normalsize{$1.48^{+0.13}_{-0.22}$}\\

 \normalsize{gabs}  & \normalsize{$\mathrm{E_{Cycl}} $[keV]} & \normalsize{$43.6^{+3.0}_{-2.2}$} & \normalsize{$43.39^{+1.41}_{-1.34}$}  & \normalsize{$39.71^{+2.16}_{-1.71}$} & \normalsize{$41.14^{+3.02}_{-1.84}$}\\

 & \normalsize{$\sigma_{\mathrm{Cycl} }$[keV]} & \normalsize{$5.8^{+2.1}_{-1.9}$} & \normalsize{5.81 (fixed)} & \normalsize{$2.17^{+2.38}_{-2.02}$}  & \normalsize{$3.46^{+2.91}_{-2.21}$}\\

  & \normalsize{$\tau_{\mathrm{Cycl} }$} & \normalsize{$0.76^{+0.65}_{-0.31}$} & \normalsize{$0.74^{+0.18}_{-0.15}$} & \normalsize{$0.59^{+15.18}_{-0.43}$} & \normalsize{$0.58^{+2.34}_{-0.40}$}\\

\normalsize{Unabsorbed flux$^{*}$} &   \normalsize{0.5-80 ~keV} & \normalsize{${2.50^{+0.04}_{-0.04}}$} & \normalsize{${2.42^{+0.02}_{-0.02}}$} & \normalsize{${2.11^{+0.03}_{-0.02}}$} & \normalsize{${2.44^{+0.02}_{-0.02}}$}\\

\normalsize{Unabsorbed luminosity$^{\#}$} &   \normalsize{0.5-80 ~keV} & \normalsize{${2.70^{+0.04}_{-0.04}}$} & \normalsize{${2.61^{+0.03}_{-0.02}}$} & \normalsize{${2.28^{+0.03}_{-0.02}}$} & \normalsize{${2.63^{+0.02}_{-0.02}}$}\\

\hline 
 & \normalsize{$\chi^2$/d.o.f} & \normalsize{777.62/650}& \normalsize{758.86/650} & \normalsize{644.34/376} & \normalsize{741.02/648}\\
 & \normalsize{${\chi^2}_{red}$} & \normalsize{1.2}& \normalsize{1.17} & \normalsize{1.71} & \normalsize{1.14}\\

\hline 
\hline 

\end{tabular}
\begin{flushleft}
 \normalsize{$^*$Flux in units of $10^{-9}$ \,erg\,cm$^{-2}$\,s$^{-1}$.}\\
\normalsize{$^\#$Luminosity  (for a distance of 9.5 kpc) in units of $10^{37}$\,erg\,s$^{-1}$.}\\
\end{flushleft}

\label{table:nonlin} 
\end{table*}

\begin{table}
\caption{\normalsize{SXT and LAXPC simultaneous spectral-fit results for XTE J1946+274 using data from 2021 \textit{AstroSat} observations of XTE J1946+274.}} 
\label{t3}
\begin{tabular}{lcrrrr} 
\hline\hline 
\normalsize{Model} & \normalsize{Parameter} & \normalsize{HIGHECUT} & \normalsize{NEWHCUT} & \normalsize{NPEX} & \normalsize{FDCUT}\\  
\hline 
\normalsize{constant} & \normalsize{LAXPC spectrum}& \normalsize{1.0 (fixed)}& \normalsize{1.0 (fixed)}& \normalsize{1.0 (fixed)} & \normalsize{1.0 (fixed)}\\

\normalsize{constant} & \normalsize{SXT spectrum} & \normalsize{$\sim 0.89$}& \normalsize{$\sim 0.9$} & \normalsize{$\sim 0.85$} & \normalsize{$\sim 0.92$}\\

\normalsize{gain offset} & \normalsize{SXT spectrum} & \normalsize{$\sim 107 ~eV$}& \normalsize{$\sim 107 ~eV$} & \normalsize{$\sim 112 ~eV$} & \normalsize{$\sim 96 ~eV$}\\

\normalsize{tbabs} & \normalsize{$\mathrm{N_{H}[10^{22}}$ \,cm$^{-2}]$} & \normalsize{$3.00^{+0.36}_{-0.25}$}& \normalsize{$2.72^{+0.28}_{-0.22}$}& \normalsize{$3.14^{+0.28}_{-0.27}$} & \normalsize{$2.82^{+0.23}_{-0.23}$}\\

\normalsize{powerlaw}    &  \normalsize{$\Gamma$} & \normalsize{$0.85^{+0.10}_{-0.06}$} & \normalsize{$0.86^{+0.03}_{-0.03}$} & \normalsize{$0.65^{+0.02}_{-0.02}$} & \normalsize{$0.90^{+0.04}_{-0.03}$}\\

\normalsize{highecut}  &   \normalsize{$\mathrm{E}_{\mathrm{cut}}$[keV]} & \normalsize{$3.99^{+0.24}_{-0.40}$} & \normalsize{$2.44^{+0.34}_{-0.48}$} & \normalsize{$9.20^{+0.22}_{-0.21}$} & \normalsize{$8.00^{+0.19}_{-0.20}$}\\

 & \normalsize{$\mathrm{E}_{\mathrm{fold}}$[keV]} & \normalsize{$20.63^{+0.53}_{-0.51}$} & \normalsize{$26.42^{+2.17}_{-1.05}$} & \normalsize{-} & \normalsize{$17.67^{+0.32}_{-0.31}$} \\

\normalsize{Gaussian}  & \normalsize{$\mathrm{E(Fe} ~\mathrm{line})$[keV]} & \normalsize{6.4 (fixed)} & \normalsize{6.4 (fixed)} &\normalsize{6.4 (fixed)} & \normalsize{6.4 (fixed)}\\

 & \normalsize{$\sigma(\mathrm{Fe} ~\mathrm{line})$[keV]} & \normalsize{$0.21^{+0.57}_{-1.30}$} & \normalsize{$1.40^{+0.09}_{-0.09}$} & \normalsize{0.21 (fixed)} & \normalsize{$1.56^{+0.09}_{-0.07}$}\\

 \normalsize{gabs}  & \normalsize{$\mathrm{E_{Cycl}} $[keV]} & \normalsize{$43.70^{+2.80}_{-2.04}$} & \normalsize{$43.06^{+2.14}_{-1.52}$}  & \normalsize{$41.82^{+1.58}_{-1.50}$} & \normalsize{$41.29^{+2.13}_{-1.56}$}\\

 & \normalsize{$\sigma_{\mathrm{Cycl} }$[keV]} & \normalsize{$9.40^{+1.56}_{-1.14}$} & \normalsize{$10.03^{+1.54}_{-0.79}$} & \normalsize{9.36 (fixed)}  & \normalsize{$9.98^{+0.97}_{-0.71}$}\\

  & \normalsize{$\tau_{\mathrm{Cycl} }$} & \normalsize{$0.62^{+0.28}_{-0.18}$} & \normalsize{$1.01^{+0.43}_{-0.27}$} & \normalsize{$1.21^{+0.46}_{-0.34}$} & \normalsize{$0.70^{+0.19}_{-0.14}$}\\

\normalsize{Unabsorbed flux$^{*}$} &   \normalsize{0.5-80 ~keV} & \normalsize{${2.08^{+0.04}_{-0.02}}$} & \normalsize{${2.17^{+0.01}_{-0.02}}$} & \normalsize{${2.11^{+0.03}_{-0.02}}$} & \normalsize{${2.09^{+0.02}_{-0.01}}$}\\

\normalsize{Unabsorbed luminosity$^{\#}$} &   \normalsize{0.5-80 ~keV} & \normalsize{${2.25^{+0.02}_{-0.03}}$} & \normalsize{${2.34^{+0.01}_{-0.02}}$} & \normalsize{${2.28^{+0.03}_{-0.02}}$} & \normalsize{${2.26^{+0.01}_{-0.01}}$}\\

\hline 
 & \normalsize{$\chi^2$/d.o.f} & \normalsize{420.48/377}& \normalsize{534.08/377} & \normalsize{644.34/376} & \normalsize{500.33/377}\\
 & \normalsize{${\chi^2}_{red}$} & \normalsize{1.12}& \normalsize{1.42} & \normalsize{1.71} & \normalsize{1.33}\\

\hline 
\hline 

\end{tabular}
\begin{flushleft}
 \normalsize{$^*$Flux in units of $10^{-9}$ \,erg\,cm$^{-2}$\,s$^{-1}$.}\\
\normalsize{$^\#$Luminosity  (for a distance of 9.5 kpc) in units of $10^{37}$\,erg\,s$^{-1}$.}\\
\end{flushleft}

\label{table:nonlin} 
\end{table}

\subsubsection{Spectral studies using 2021 observations}
\begin{figure}
       \centering
      \includegraphics[width=0.8\linewidth]{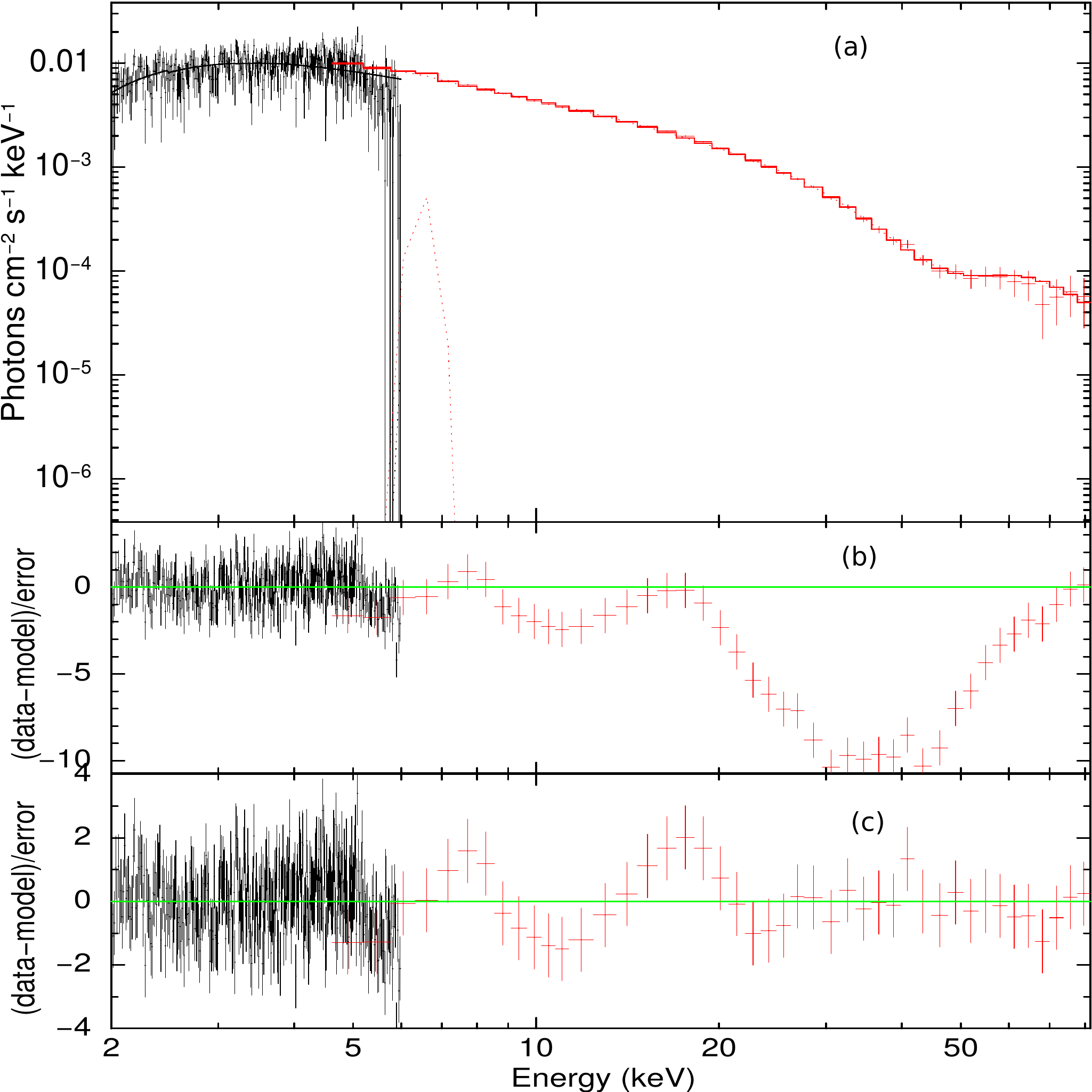}
    \caption{ (a) Simultaneous fitted SXT and LAXPC spectrum for the 2021 outburst using the power law with high energy cut-off model. The solid line shows the best-fitting model along with the spectral data. (b) Prominent absorption residuals around 43 keV are clearly seen in the panel without fitting for any cyclotron line absorption features. (c) A cyclotron line around 43 keV is fitted to the model and residuals between the data and the model are shown in the panel. The significance of detection of the cyclotron line is more than 3$\sigma$.}
    \label{f11}
\end{figure}

We have performed broad-band (2-80 keV) spectral fitting of the SXT
and LAXPC20 spectra for the 2021 observation (Fig. $\ref{f11}$). The broad-band spectral fitting is confined to 2-80 keV energy range owing to relatively poorer SNR of the spectra compared to the 2018 \textit{AstroSat} observations.
We have used the same models for spectral fitting as used earlier for the 2018 observations. The procedures of applying gain correction and accounting for cross calibration differences between the SXT and LAXPC20 spectra are already described in the previous section.
The energy of the iron emission line was frozen at 6.4 keV during the spectral fitting. The results of the broad-band spectral fitting for the 2021 observations are given in table \ref{t3}.
The spectrum of the pulsar during the 2021 outburst is found to be relatively soft ($\Gamma \sim 0.9$) compared to that of the 2018 outburst. The absorption $\mathrm{N_H}$ is deduced to be $\sim 3\times 10^{22}$ \,cm$^{-2}$ from the four spectral models used to describe the broad-band spectra. A Gaussian optical depth proﬁle \textit{gabs} was used to fit deviations in the fitted spectrum around 43 keV (Fig. \ref{f11}). The unabsorbed X-ray flux of XTE J1946+274 in the 0.5-80 keV energy interval for the
2021 outburst is estimated to be $2.1 \times 10^{-9}$ \,erg\,cm$^{-2}$\,s$^{-1}$ which implies X-ray luminosity to be $2.3 \times 10^{37}$\,erg\,s$^{-1}$ for a distance of 9.5 kpc for the source.

Table ${\ref{t5}}$ summarizes characteristics of the cyclotron lines detected during different outbursts in this pulsar since its discovery in 1998. It is observed that the cyclotron line energy lies in the range of about 35-43 keV, except for the report of detection of a CRSF at $\sim$25 keV in the \textit{RXTE} observations during the 2010 outburst (\citealt{muller2012reawakening}). The 6 \textit{BeppoSAX} observations of the
2010 outburst in 2010 October-November listed in Table ${\ref{t5}}$, lead to detection of cyclotron line at
energy ranging from 35 keV to 40.8 keV. This suggests that either the line energy varies with the orbital phase or source intensity. The two \textit{Suzaku} results, one based on the analysis by \cite{maitra2013pulse} lead to energy of CRSF as 38.3 keV while analysis of the same \textit{Suzaku} data by \cite{marcu2015transient} yields a line energy of 35.16 keV. This suggests that the deduction of the line energy is dependent on the fitted model. This also casts doubt on the reality of the differences in the energy of the line inferred from different observations.

\begin{table}
\caption{\normalsize{List showing characteristics of cyclotron lines detected during outbursts in the X-ray pulsar XTE J1946+274.}} 
\label{t5}
\centering 
\setlength{\tabcolsep}{1pt}
\small
\begin{tabular}{c|c|c|c|c|c|c|c|c} 
\hline\hline 
\normalsize{Observation time} & \normalsize{Satellite} & \normalsize{$\mathrm{E_{Cycl}} $[keV]} & \normalsize{$\sigma_{\mathrm{Cycl} }$[keV]} & \normalsize{$\tau_{\mathrm{Cycl}}$} & \normalsize{$\mathrm{Flux^{*}}$} & \normalsize{$\mathrm{{L_X}^{\#}}$} & \normalsize{E[keV]}& \normalsize{Ref.}\\  
\hline
 \normalsize{September 16-October 8 1998} & \normalsize{\textit{RXTE}} & \normalsize{$36.2^{+0.5}_{-0.7}$} & \normalsize{$3.37^{+0.92}_{-0.75}$} & \normalsize{$0.33^{+0.07}_{-0.06}$} & \normalsize{5.5} & \normalsize{5.94} & \normalsize{2-10} & \normalsize{1}\\

 \normalsize{October 9 1998}  & \normalsize{\textit{BeppoSAX}} & \normalsize{$38.34^{+1.45}_{-1.3}$} & \normalsize{$4.55^{+1.35}_{-1.21}$} & \normalsize{$0.3^{+0.1}_{-0.1}$} &  \normalsize{4.42} & \normalsize{4.48} & \normalsize{0.1-120} & \normalsize{2}\\

 \normalsize{October 22 1998}  & \normalsize{\textit{BeppoSAX}} & \normalsize{$40.79^{+2.2}_{-1.92}$} & \normalsize{$4.0 \rm{(fixed)}$} & \normalsize{$0.3^{+0.1}_{-0.2}$} & \normalsize{3.15} & \normalsize{3.4} & \normalsize{0.1-120} & \normalsize{2}\\

 \normalsize{November 8 1998}  & \normalsize{\textit{BeppoSAX}} & \normalsize{$36 \rm{(fixed)}$} & \normalsize{$4.0 \rm{(fixed)}$} & \normalsize{$0.12^{+0.01}_{-0.01}$} & \normalsize{1.37} & \normalsize{1.48} & \normalsize{0.1-120} & \normalsize{2}\\

 \normalsize{November 12 1998}  & \normalsize{\textit{BeppoSAX}} & \normalsize{$38.33^{+2.76}_{-1.95}$} & \normalsize{$0.9^{+2.8}_{-0.8}$} & \normalsize{$1.1^{+0.9}_{-0.9}$} & \normalsize{1.16} & \normalsize{1.25} & \normalsize{0.1-120} & \normalsize{2}\\

 \normalsize{November 22 1998}  & \normalsize{\textit{BeppoSAX}} & \normalsize{$38 \rm{(fixed)}$} & \normalsize{$4.0 \rm{(fixed)}$} & \normalsize{$0.59^{+0.01}_{-0.01}$} &\normalsize{0.63} & \normalsize{0.68} & \normalsize{0.1-120} & \normalsize{2}\\

 \normalsize{November 27 1998}  & \normalsize{\textit{BeppoSAX}} & \normalsize{$35.17^{+2.51}_{-1.78}$} & \normalsize{$0.4^{+2.9}_{-0.3}$} & \normalsize{$0.3^{+0.7}_{-0.1}$} & \normalsize{0.42} & \normalsize{0.45} & \normalsize{0.1-120} & \normalsize{2}\\

  \normalsize{June 20-30 2010}  & \normalsize{\textit{RXTE, INTEGRAL}} & \normalsize{$25.3^{+0.9}_{-1.0}$} & \normalsize{$0.65^{+1.46}_{-0.15}$} & \normalsize{$0.09^{+0.10}_{-0.07}$} & \normalsize{12.45$^\$$} & \normalsize{13.45} & \normalsize{3-80} & \normalsize{3}\\

  \normalsize{July 3-16 2010}  & \normalsize{\textit{RXTE}} & \normalsize{$25.3 \rm{(fixed)}$} & \normalsize{$0.65 \rm{(fixed)}$} & \normalsize{$0.03^{+0.08}_{-0.03}$} & \normalsize{7.9$^\$$} & \normalsize{8.53} & \normalsize{3-80} & \normalsize{3}\\

\normalsize{October 11-13 2010}  & \normalsize{\textit{Suzaku}} & \normalsize{$38.30^{+1.63}_{-1.36}$} & \normalsize{$9.61^{+3.69}_{-3.06}$} & \normalsize{$1.72^{+0.41}_{-0.28}$} & \normalsize{0.53} & \normalsize{0.57} & \normalsize{0.3-70} & \normalsize{4}\\

\normalsize{October 11-13 2010}  & \normalsize{\textit{Suzaku}} & \normalsize{$35.16^{+1.5}_{-1.3}$} & \normalsize{$2 \rm{(fixed)}$} & \normalsize{$0.48^{+0.30}_{-0.26}$} & \normalsize{0.34} & \normalsize{0.37} & \normalsize{10-40} & \normalsize{5}\\

 \normalsize{November 24-28 2010}  & \normalsize{\textit{RXTE, Swift}} & \normalsize{$25.3 \rm{(fixed)}$} & \normalsize{$0.65 \rm{(fixed)}$} & \normalsize{$\leq 0.23$} & \normalsize{5.78$^\$$} & \normalsize{6.24} & \normalsize{3-80} & \normalsize{3}\\

 \normalsize{November 30-December 3 2010}  & \normalsize{\textit{RXTE, INTEGRAL, Swift}} & \normalsize{$25.3 \rm{(fixed)}$} & \normalsize{$0.65 \rm{(fixed)}$} & \normalsize{$\leq 0.40$} & \normalsize{5.03$^\$$} & \normalsize{5.43} & \normalsize{3-80} & \normalsize{3}\\

\normalsize{June 24 2018} & \normalsize{\textit{NuSTAR}} & \normalsize{$37.49^{+0.70}_{-0.64}$} & \normalsize{$8.59^{+0.69}_{-0.61}$} & \normalsize{$0.567^{+0.011}_{-0.011}$} & \normalsize{2.603} & \normalsize{2.8} & \normalsize{3-79} &\normalsize{6}\\

\normalsize{June 24 2018} & \normalsize{\textit{NuSTAR}} & \normalsize{$37.81^{+0.75}_{-0.73}$} & \normalsize{$4.53^{+0.59}_{-0.55}$} & \normalsize{$0.248^{+0.003}_{-0.003}$} & \normalsize{2.606} & \normalsize{2.8} & \normalsize{3-79} & \normalsize{6}\\

\normalsize{June 9-10 2018} & \normalsize{\textit{AstroSat}} & \normalsize{$43.60^{+3.0}_{-2.2}$} & \normalsize{$5.8^{+2.1}_{-1.9}$} & \normalsize{$0.80^{+0.88}_{-0.37}$} & \normalsize{2.1-2.5} & \normalsize{2.3-2.7} & \normalsize{3-80} & \normalsize{7}\\

\normalsize{October 3-6 2021} & \normalsize{\textit{AstroSat}} & \normalsize{$43.6^{+2.9}_{-2.9}$} & \normalsize{$9.3^{+2.6}_{-2.1}$} & \normalsize{$0.60^{+0.50}_{-0.33}$} & \normalsize{2.1} & \normalsize{2.3} & \normalsize{3-80} & \normalsize{7}\\
\hline 
\hline 

\end{tabular}
\label{table:nonlin}
\\
\begin{flushleft}
\textit{Notes.}
\normalsize{$^{*}$Flux in units of $10^{-9}$ \,erg\,cm$^{-2}$\,s$^{-1}$.}\\
\normalsize{$^{\#}$Luminosity  (for a distance of 9.5 kpc) in units of $10^{37}$\,erg\,s$^{-1}$.}\\
\normalsize{$^{\$}$Flux obtained in 3-80 keV energy range using  WebPIMMS (\url{https://heasarc.gsfc.nasa.gov/cgi-bin/Tools/w3pimms/w3pimms.pl}).}\\
(1) \cite{heindl2001discovery} (2) \cite{doroshenko2017bepposax} (3) \cite{muller2012reawakening} (4) \cite{maitra2013pulse} (5) \cite{marcu2015transient} (6) \cite{gorban2021study} (7) this work.\\
\end{flushleft}
\end{table}

\subsection{Accretion regimes in XTE J1946+274}
The pulse profiles are manifestation of the complex plasma dynamics in the vicinity of the neutron star. The accreted material channelled along the magnetic  field lines of the neutron star hits the surface of the neutron star and forms a mound which emits black-body thermal radiation  (\citealt{becker2007thermal}). As more accreted material falls near the surface of the neutron star, the thermal radiation from the mound increases and forms a
shock at the  interface between the mound and the infalling matter. This shock front rises away from the neutron star surface and forms an accretion column underneath (\citealt{basko1976}).  The formation and growth of these accretion columns are governed by the  tug of war between gravity and radiation pressure.
Depending on the X-ray luminosity, two regimes of accretion viz. sub-critical  and super-critical accretion takes place. The threshold luminosity which separates these two accretion modes is known as the critical luminosity ($L_{X,crit}$, \cite{becker2012spectral}). The accretion stream is decelerated by thermal electrons through Coulomb forces in the sub-critical accretion mode ($L_X < L_{X,crit}$) and radiation is emitted along the local magnetic field lines giving rise to a \textquoteleft pencil-beam\textquoteright pattern. In the other accretion regime ($L_X > L_{X,crit}$), formation of an accretion column takes place and maximum radiation is emitted perpendicular to the magnetic dipole axis resulting in a \textquoteleft fan-beam\textquoteright pattern. The critical luminosity is given by (\citealt{becker2012spectral}),
\begin{equation}
\begin{aligned}
L_{X,crit} = 1.49 \times 10^{37} \, {\rm erg\ s^{-1}}
& \left(\Lambda \over 0.1\right)^{-7/5} w^{-28/15} \\
\times \ \left(M_* \over 1.4 M_\odot\right)^{29/30}
& \left(R_* \over 10\,{\rm km}\right)^{1/10}
\left(B_* \over 10^{12}{\rm G}\right)^{16/15}
\ ,
\end{aligned}
\label{eq32}
\end{equation}

where $\Lambda$ is a constant parameter ($\Lambda=1$ for spherical accretion and $\Lambda < 1$ for disk accretion, (\citealt{becker2012spectral})), $w$ depends on the spectrum in the accretion column and  $w=1$ for the dominant bremsstrahlung emission inside the accretion column (\citealt{becker2007thermal}) and  $M_*$, $R_*$ and $B_*$ are the mass, radius and magnetic field of the neutron star respectively. Using typical neutron star parameters $M_*=1.4\,M_\odot$, $R_*=10\,$km, $\Lambda=0.1$, and $w=1$, $L_{X,crit} = 1.49 \times 10^{37} \, {\rm erg\ s^{-1}} B_{12}^{16/15}$, where $B_{12}$ is the surface magnetic field strength (\citealt{becker2012spectral}). For surface magnetic field of about $3.1 \times 10^{12}$G in XTE J1946+274 (\citealt{heindl2001discovery}), the estimated critical luminosity is $L_{X,crit} \sim 5\times 10^{37}{\rm erg\ s^{-1}}$.

An updated version of the critical luminosity was calculated by \cite{mushtukov2015critical} given by,

\begin{equation}
\begin{aligned}
L^*_{X,crit} \approx \frac{c}{\kappa_{eff}} \pi d \frac{GM_*}{R_*} \approx 3.7 \times 10^{36} \left(\kappa_T \over \kappa_{eff} \right)
& \frac{d_5}{R_{*6}} ~m \, {\rm erg\ s^{-1}},
\end{aligned}
\label{eq33}
\end{equation}

where $\kappa_{eff}$ is the effective opacity, d ($d\ll R_*$) is the diameter of a bright axisymmetric spot created on the neutron star surface due to heating by the accreting matter, $\kappa_T$ $\approx$ 0.34 $\rm{cm^2\ g^{-1}}$ is the Thomson scattering opacity for solar composition material, $m=\frac{M}{M_\odot}$ (\citealt{mushtukov2015critical}).

In the sub-critical emission regime, the accreted matter can be decelerated by the radiation shock followed by Coulomb interactions for moderate sub-critical luminosities ($L_X \lesssim L_{X,crit}$) while at even lower luminosities ($L_X \lesssim L_{X,Coul}$) the radiation shock and Coulomb braking mechanism disappear and the accreted matter falls through a gas-mediated shock before hitting the neutron star surface (\citealt{langer1982low, becker2012spectral}). The Coulomb luminosity $L_{X,Coul}$is given by (\citealt{becker2012spectral}),

\begin{equation}
\begin{aligned}
L_{X,Coul} = 1.17 \times 10^{37} \, {\rm erg\ s^{-1}}
& \left(\Lambda \over 0.1\right)^{-7/12}
\left(\tau_* \over 20\right)^{7/12}
\left(M_* \over 1.4 M_\odot\right)^{11/8} \\
\times \
& \left(R_* \over 10\,{\rm km}\right)^{-13/24}
\left(B_* \over 10^{12}{\rm G}\right)^{-1/3}
\,
\end{aligned}
\label{eq54}
\end{equation}

where $\tau_*\sim20$ is the Thomson depth required for the accreted gas to be effectively stopped via Coulomb interactions (\citealt{becker2012spectral}). Using typical neutron star parameters $\Lambda=0.1$, $\tau_*=20$, $M_*=1.4\,M_\odot$, $R_*=10\,$km, and $B_* \sim 3.1 \times 10^{12}$G for XTE 1946+274 we obtain $L_{X,Coul} \sim 8\times 10^{36}{\rm erg\ s^{-1}}$. Thus, the accretion luminosities during the 2018 and 2021 outburst of this pulsar are $\sim 2.7\times 10^{37}{\rm erg\ s^{-1}}$ and $\sim 2.3\times 10^{37}{\rm erg\ s^{-1}}$ which lie in the sub-critical regime ($L_X < L_{X,crit}$) such that $L_X > L_{X,Coul}$. In this accretion regime, the emission beam pattern can be described by a combination of a \textquoteleft pencil-beam\textquoteright and a \textquoteleft fan-beam\textquoteright (ref. schematic illustration shown in Fig. 1 of \cite{becker2012spectral}). It should be noted that XTE J1946+274 has been observed during outbursts by and large in this sub-critical accretion regime and hence the pulse profiles and likely the underlying beam pattern do not show any signiﬁcant temporal evolution. Several pointed observations of the pulsar during different luminosity levels in future outbursts of this pulsar can be helpful to probe luminosity dependent profile evolution in this pulsar in more detail.

\subsection{Exploring dependence of peak separation on X-ray luminosity}
\begin{figure}
       \centering
      \includegraphics[width=\linewidth]{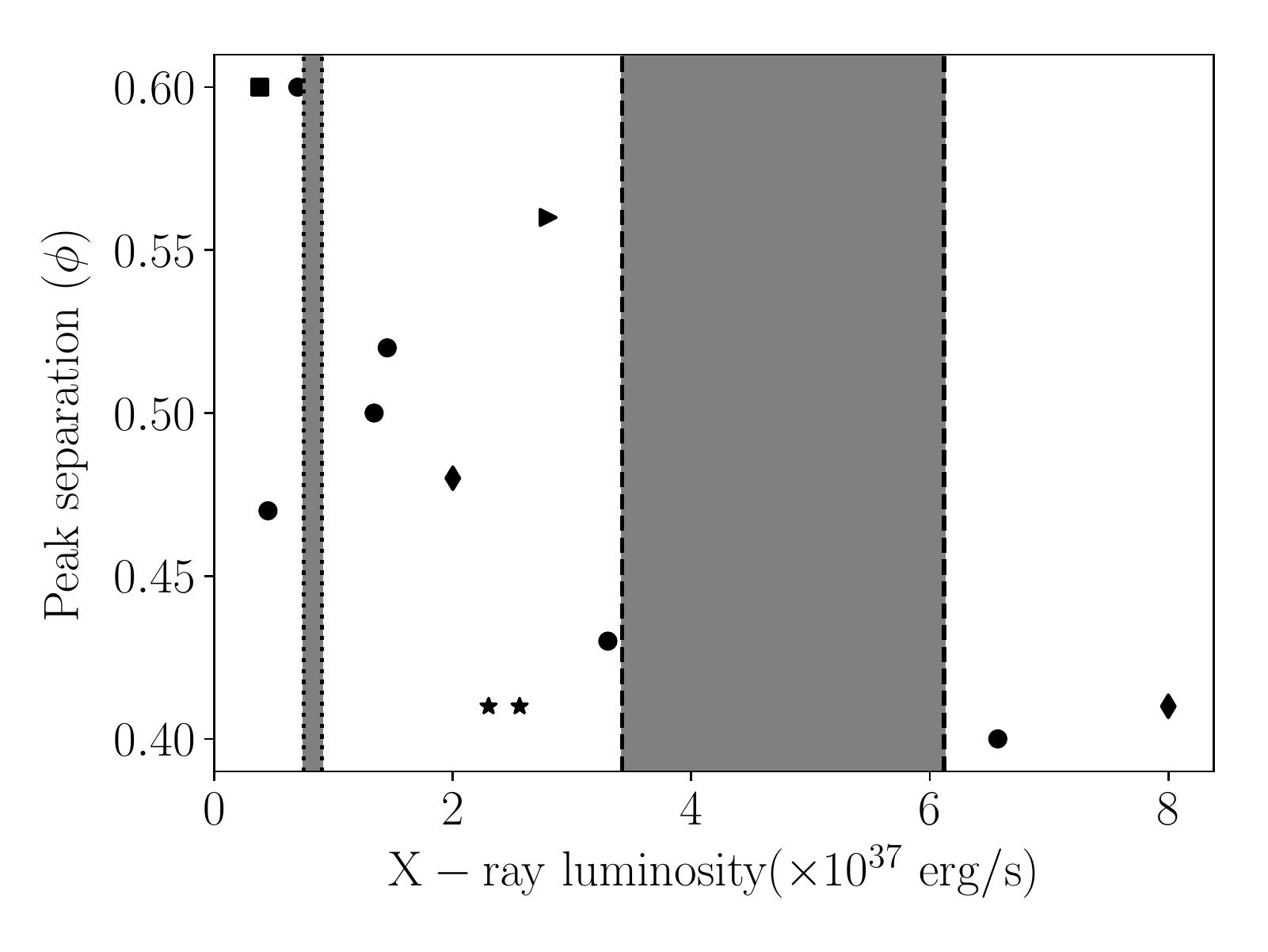}
    \caption{Peak separation vs X-ray luminosity for pulse profiles used in this study and those obtained from literature ($\blacklozenge$: \cite{wilson2003xte}, $\bullet$: \cite{doroshenko2017bepposax}, $\blacksquare$: \cite{maitra2013pulse}, $\blacktriangleright$: \cite{gorban2021study}, $\bigstar$: this work). 
    The grey-shaded regions bordered with dotted and dashed lines show range of possible values for $L_{X,Coul}$ and $L_{X,crit}$ respectively using different magnetic field values inferred from cyclotron line energies tabulated in Table \ref{t5}.}
    \label{f11a}
\end{figure}

We compute the relative separation between the two peaks in the pulse profiles (the two peaks are clearly detected for energies below 20 keV) and explore its possible dependence on X-ray luminosity using the 2018 and the 2021 \textit{AstroSat}/LAXPC observations and from the pulse profiles reported in literature (\citealt{wilson2003xte,doroshenko2017bepposax,gorban2021study}). It should be noted that although the pulse profiles evolve with energy but the relative separation between the two peaks in the pulse profiles remain nearly constant with energy for energies below 20 keV. Above 20 keV the second peak in the pulse profile usually becomes flat and it becomes difficult to estimate the relative separation between the two peaks in the pulse profiles. We observe a
possible anti-correlation between the separation of the peaks in the pulse profiles and the X-ray luminosity as shown in Fig. \ref{f11a}. The vertical dotted lines shown in Fig. \ref{f11a} show the estimated Coulomb luminosity ($L_{X,Coul}\sim 8\times 10^{36}{\rm erg\ s^{-1}}$) and critical luminosity ($L_{X,crit} \sim 5\times 10^{37}{\rm erg\ s^{-1}}$) of the pulsar. Indication of correlation between the separation of the peaks in the pulse profiles and the X-ray luminosity has been observed in another Be X-ray binary pulsar SXP 1062 by \cite{cappallo2020geometry}. We note this trend is opposite to what is noticed in XTE J1946+274. It has been suggested that the critical luminosity marks the transition in the beam profile shape from \textquotedblleft pencil-beam \textquotedblright to \textquotedblleft fan-beam \textquotedblright (\citealt{becker2012spectral}). Changes in the beam pattern of the pulsar with luminosity would manifest as changes in the pulse shape and separation between the peaks in the folded profiles. It is expected that as the pulse profile becomes dominated by the \textquotedblleft fan-beam \textquotedblright pattern with concomitant formation of an accretion column with increasing luminosity ($L_{X}\gtrsim L_{X,crit}$,\cite{becker2012spectral}), photons are primarily emitted from the side of the accretion column and so the separation between the peaks in the folded profile should also increase.

\subsection{Exploring dependence of cyclotron line energy on X-ray luminosity}
\begin{figure}
       \centering
      \includegraphics[width=\linewidth]{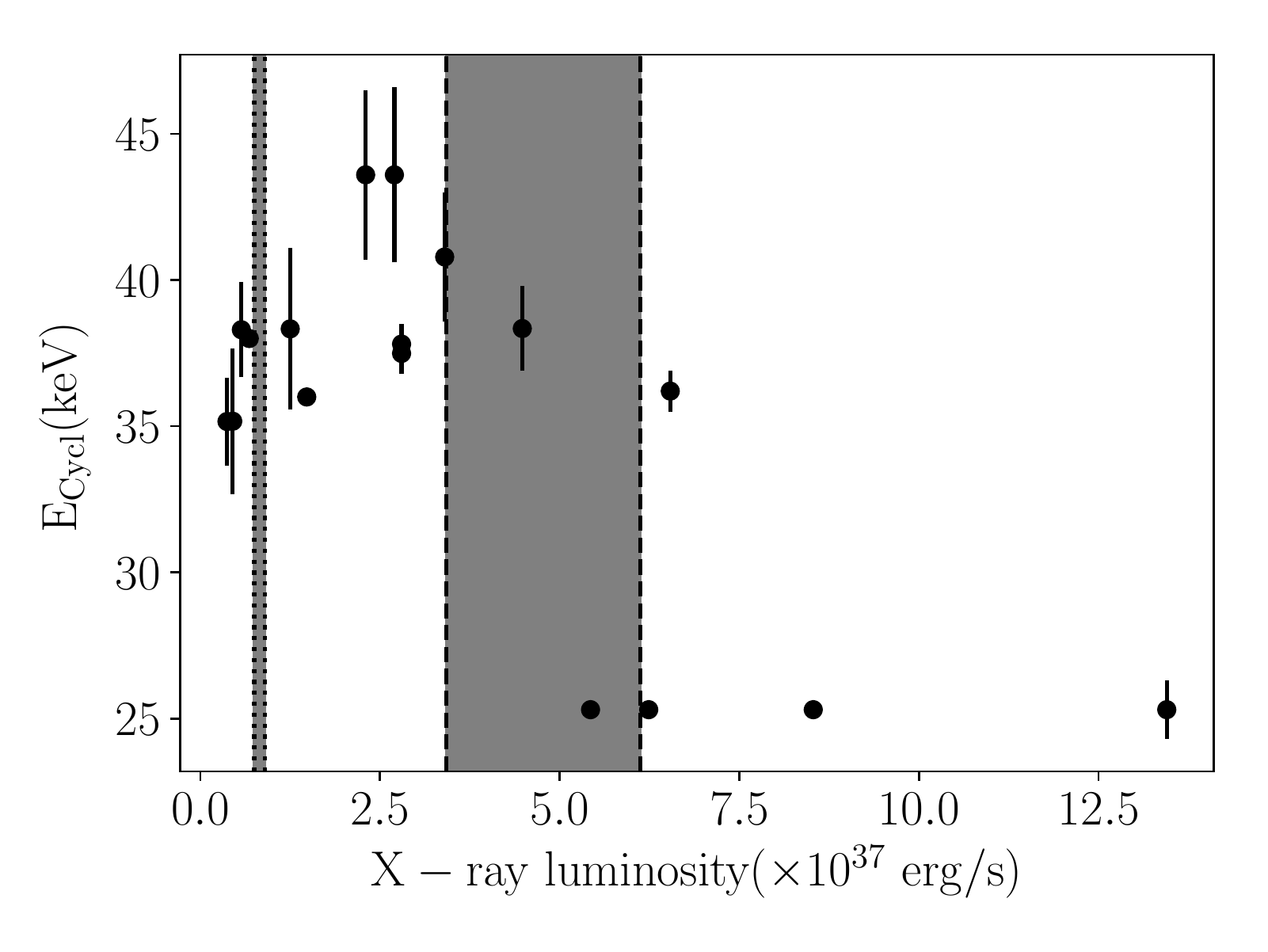}
    \caption{Dependence of cyclotron line energy on X-ray luminosity for XTE
J1946+274. The grey-shaded regions bordered with dotted and dashed lines show range of
possible values for $L_{X,Coul}$ and $L_{X,crit}$ respectively using different magnetic field values inferred from cyclotron line energies tabulated in Table \ref{t5}.}
    \label{f11b}
\end{figure}

We investigate variations of cyclotron line energy with X-ray luminosity in Fig. \ref{f11b}. There is suggestion of positive correlation of cyclotron line energy with luminosity until around $L_{X} \sim 3\times 10^{37}{\rm erg\ s^{-1}}$ after which negative correlation is observed.
Negative correlation of cyclotron line energies with luminosities has been detected in transient Be/X-ray binaries V 0332+53 (\citealt{makishima1990discovery,mihara1995observational,tsygankov2006v0332}) and SMC X-2 (\citealt{jaisawal2016detection}). A positive correlation of cyclotron line energy with luminosity has been detected in six accreting pulsars
Vela X-1 (\citealt{fuerst2013nustar,la2016swift}), A 0535+26 (\citealt{klochkov2011pulse,sartore2015integral}), GX 304-1 (\citealt{yamamoto2011discovery,malacaria2015luminosity,rothschild2017discovery}), Cep X-4 (\citealt{vybornov2017luminosity}), and 4U 1626.6-5156 (\citealt{decesar2012x}).

Interestingly, in the transient Be binary V0332+53, a negative correlation between cyclotron line energy and luminosity was detected at high luminosity which reversed to a positive correlation at lower luminosity during end phases of an outburst (\citealt{doroshenko2017luminosity,vybornov2018changes}). The reversal in the correlation between the cyclotron line energy and luminosity was separated by the critical luminosity for this source and was suggested to be caused due to variation in the emission region geometry (\citealt{doroshenko2017luminosity}). However, we note that in case of XTE
J1946+274, the possible reversal in the correlation between the cyclotron line energy and luminosity is $\sim 3\times 10^{37}{\rm erg\ s^{-1}}$ which is slightly less than the estimated critical luminosity of $\sim 5\times 10^{37}{\rm erg\ s^{-1}}$ for this source.
It is likely that the possible reversal in the correlation between the cyclotron line energy and luminosity in XTE J1946+274 is also due to changes in emission geometry of the pulsar.
Further luminosity resolved spectral investigations of XTE
J1946+274 during future outbursts are required to confirm the observed behaviour between cyclotron line energy and luminosity.

\begin{figure}
        \centering
        \includegraphics[width=\linewidth]{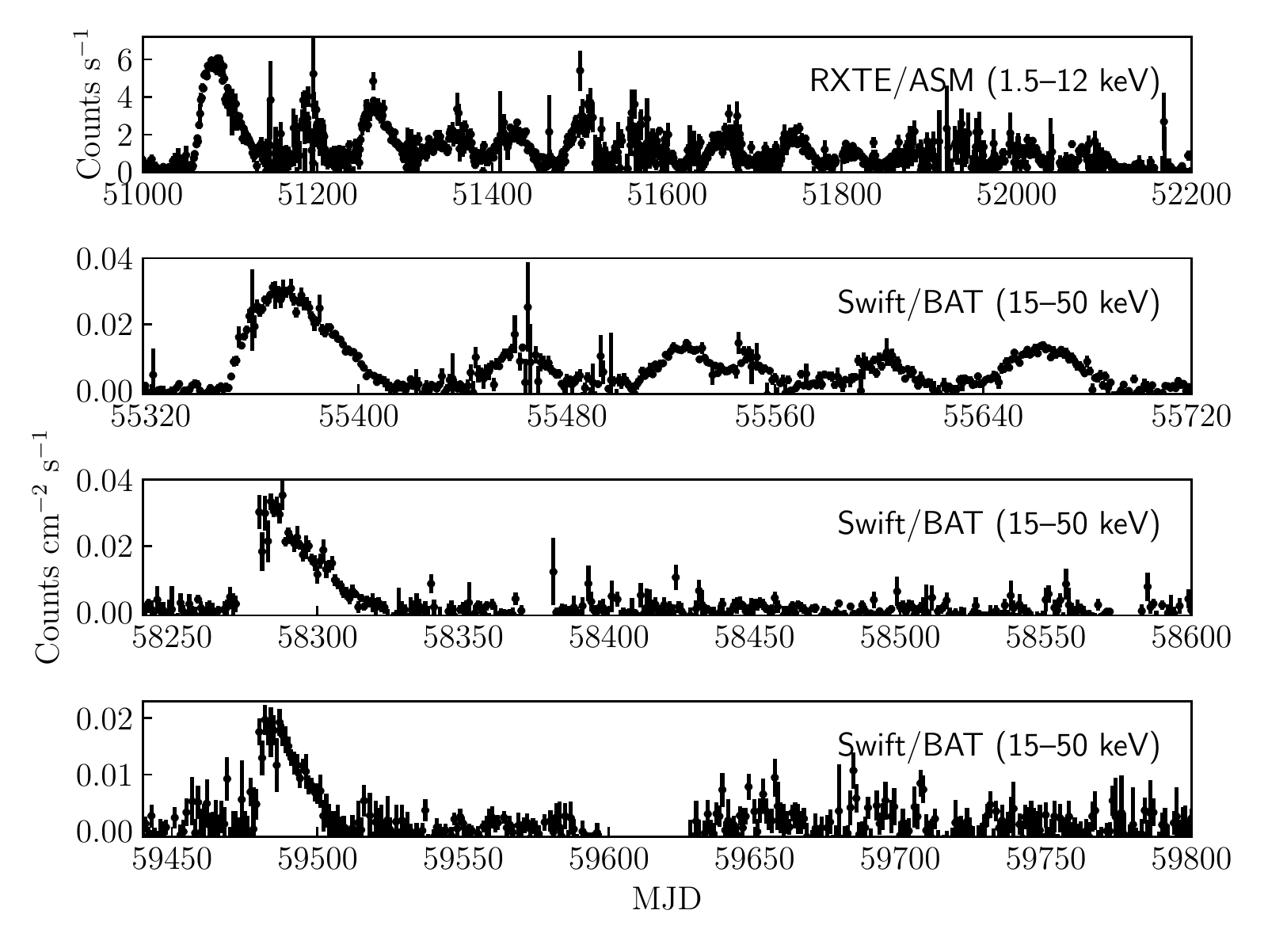}
    \caption{\textit{RXTE}/ASM one day averaged light curve of the 1998 outburst of XTE
J1946+274 shown in the top panel. \textit{Swift}/BAT one day averaged light curve of the 2010, 2018 and 2021 outbursts of XTE J1946+274 are shown in the second, third and the fourth panels (top to bottom).}
    \label{f12}
\end{figure}

\subsection{Variety of outbursts in XTE J1946+274}
XTE J1946+274 is an enigmatic Be/XRB which has undergone four outbursts since its discovery more than two decades ago in 1998. Fig. \ref{f12} shows X-ray monitoring observations of the outbursts detected in this pulsar using the \textit{RXTE}/ASM during 1998-2001 and using the \textit{Swift}/BAT during 2010-2011, 2018 and 2021 outbursts. It is observed that the pulsar has undergone outbursts which differ in duration, outburst pattern, number of outbursts during multiple outbursts in 1998 and 2010 and brightness of outbursts. XTE J1946+274 remained active from 1998 September-2001 July, undergoing 13 outbursts that were not locked in orbital phase (\citealt{wilson2003xte}).

 Intriguingly, the alternate outbursts were slightly brighter during this unusual outburst episode of the pulsar and the X-ray ﬂux did not drop significantly between the outbursts suggesting that accretion was not quenched during this active period.
 It is interesting to note that the low intensity outbursts during the 2010 extended outburst of the source had almost half of the peak intensity
of the initial giant outburst as also seen in the 1998 outburst.
The remarkable similarity in the outburst pattern during the 1998 and 2010 outbursts suggests a clock-like mechanism in the underlying phenomena driving outbursts in this Be/XRB.

The duration and outburst pattern of the 2018 and 2021 outbursts were remarkably different compared to the earlier 1998 and 2010 outbursts, as only a single outburst lasting for about a few weeks was detected during their active phase. Intriguingly, these outbursts were also not phase locked similar to the previous outbursts.
The peak intensities of the 2018 and 2021 outbursts were about 0.04 $\rm{counts\,s^{-1}}$ and 0.02 $\rm{counts\,s^{-1}}$ respectively from \textit{Swift}/BAT observations. Interestingly, the peak intensities of the 2018 and 2021 outbursts were comparable to the peak intensity of the giant outburst and low intensity outbursts in 2010 respectively. The outbursts detected in this pulsar have different decay profiles which may be  classified in two broad categories: (I) a single  giant Type II outburst like those detected in the 2018 and 2021 episodes reported in this paper. (II)  in the second category the giant outburst is followed by several low intensity periodic outbursts modulated at half of the orbital period of the system but are not locked in orbital phase. The outbursts  having  a single giant outburst decay profile are also not locked to orbital phase and do not have any post cursors.

\subsection{Possible drivers of unusual outbursts}

The giant outbursts detected in XTE J1946+274 are rare and similar to the Type II outbursts detected in the other Be binaries. The peak luminosity during the 1998 outburst was about $8 \times 10^{37}$ \,erg\,s$^{-1}$ in the 2-60 keV energy range (\citealt{doroshenko2017bepposax}). The luminosity of two low intensity outbursts observed during 2001 was about $2 \times 10^{37}$ \,erg\,s$^{-1}$ in the 2-60 keV energy range (\citealt{doroshenko2017bepposax}). The inferred unabsorbed luminosities during the 2018 and 2021 outbursts of this pulsar from \textit{AstroSat} observations are about $2.7 \times 10^{37}$ \,erg\,s$^{-1}$ and $2.3 \times 10^{37}$ \,erg\,s$^{-1}$ in the 0.5-80 keV band respectively. The range of accretion powered peak luminosities detected during outbursts in this pulsar is about $2-8 \times 10^{37}$ \,erg\,s$^{-1}$ which is comparable to the typical luminosity ($10^{37}$ \,erg\,s$^{-1}$) observed during giant (Type II) outbursts in Be/XRB systems.

From investigation of H $\alpha$ line profiles during the 1998-2001 outbursts, \cite{wilson2003xte} inferred that the dynamics of the decretion disc around the Be star plays a crucial role in driving X-ray outbursts in this peculiar Be/XRB.
In addition to warping of the misaligned Be star disc (\citealt{arabaci2015detection}), it has been suggested that mass outflow from the Be star could also trigger a giant outburst in this Be binary (\citealt{wilson2003xte}). From optical/IR brightening in the Be star, \cite{arabaci2015detection} suggested that an X-ray outburst of this source was imminent in about two yr (i.e. around 2017) likely due to mass ejection from the Be star. Indeed, we detect an outburst in this Be/XRB around
2018 June which suggests that the impending outburst was delayed
by about an year. Interestingly, the next outburst in this source was
detected around 2021 September which is about 3 yr after the 2018
outburst while the gap between the 2010-2011 and 2018 outburst
was about 7 yr which is almost double of the 3 yr duration. This
suggests that the elusive phenomena which triggers outbursts in this
Be/XRB (likely ejection of mass from the Be star and factors such
as the structure and dynamics of the decretion disc) took almost
double time to reach its threshold to trigger the 2018 outburst.

The unusual outbursts detected during 1998-2001 and 2010-2011 which showed two outbursts per orbit and were not locked in orbital phase poses a conundrum. It was suggested that a decretion disc misaligned with the orbital plane of the binary system could produce two outbursts per orbit but then they should be fixed in orbital phase which is not the case in this system (\citealt{wilson2003xte}). Using 3D simulations it has been shown that the strong gravitational field of the neutron star can distort the Be disc and form strong asymmetric structures in the disc (\citealt{okazaki2011hydrodynamic}), which could likely lead to multiple X-ray outbursts in an orbit (\citealt{muller2012reawakening}). Another alternative proposition was that these outbursts could be triggered by a combination of density variations in the Be disc and orbital effects (\citealt{muller2012reawakening}). A long-term multiwavelength monitoring campaign of this pulsar is required to better probe the underlying phenomena which triggers outbursts in this Be/XRB.

\subsection{Von Zeipel-Lidov-Kozai disc oscillations mediated giant outbursts in XTE J1946+274?}
A sufficiently misaligned decretion disc around a Be star can become highly eccentric (\citealt{martin2014giant}) due to Von Zeipel-Lidov-Kozai (ZLK) oscillations (\citealt{von1910application,kozai1962secular,lidov1962evolution}). This is due to exchange between inclination and eccentricity of a misaligned orbit around one component of a binary. A Be/XRB system can exhibit ZLK disc oscillations mediated giant (Type II) outbursts provided two conditions are fulfilled (\citealt{martin2014giant}). Firstly, the decretion disc around the Be star must be inclined with respect to the orbital plane of the binary system and this inclination should exceed 39$\degree$, the
critical angle required for the onset of ZLK oscillations but it varies for thick decretion discs depending on characteristics of the disc (\citealt{lubow2017kozai}). Secondly, the
orbital period of the binary must be short enough to allow the Be star
disc to expand sufficiently far to overflow its Roche lobe during a ZLK oscillation and transfer mass onto the neutron star to trigger an outburst. XTE J1946+274 has a misaligned decretion disc which has been suggested in earlier studies (\citealt{wilson2003xte,arabaci2015detection}). However, the misalignment angle is not known and needs to be ascertained from future optical observations of this source. There are several Be/XRB systems which exhibit Type II outbursts but have a long orbital period including XTE J1946+274. 1A 0535+262, Swift J1626.6-5156 and GRO J1008-57 have orbital periods of about 111 d (\citealt{finger1996quasi}), 132 d (\citealt{baykal2010orbital}) and 250 d  (\citealt{kuhnel2013gro}) but show giant outbursts. In order for ZLK disc oscillations to drive giant outbursts in
1A 0535+262 and Swift J1626.6-5156, the decretion disc  must not be very flared while in case of GRO J1008-57, the disc aspect ratio must be close to constant with radius (\citealt{martin2014giant}). Similarly, for ZLK oscillations to operate in XTE J1946+274, the disc around the Be star should not be very flared. It has been suggested from observations in some Be/XRB systems that giant (Type II) outbursts occur when the decretion disc around the Be star becomes warped (\citealt{negueruela2001x,reig2007x,moritani2013precessing,ducci2019x}). Similarly, warping of the Be disc is surmised to trigger giant outbursts in XTE J1946+274 (\citealt{arabaci2015detection}). However, (\cite{martin2014giant}) have shown using three-dimensional smoothed particle hydrodynamics simulations of Be/XRB systems that warping of Be star disc alone is not a sufficient condition to trigger giant outbursts in these systems but it is necessary for the decretion disc to become eccentric.

In a very recent work, \cite{martin2021kozai} have shown
that ZLK disc oscillations can drive a pair of giant (Type II) outbursts separated by several orbital periods in Be/XRBs. The first
Type II outburst is triggered due to the misaligned Be disc which
undergoes ZLK oscillations while the second Type II like outburst
is driven by the eccentric accretion disc around the neutron star
which may also undergo ZLK oscillations and lead to enhanced
mass transfer onto the neutron star (\citealt{martin2021kozai}). It is suggested that the second outburst of the outburst pair is comparatively smaller than the first outburst (\citealt{martin2021kozai})
which is also observationally detected in 4U 0115+63 wherein the
second outburst is less luminous of the outburst pair (\citealt{reig2018warped}). Interestingly, in case of XTE J1946+274, we find that the
2021 outburst occurred almost 3 yr after the 2018 outburst and was
less luminous than the 2018 outburst. If we surmise that the 2018
and 2021 outbursts form an outburst pair, then XTE J1946+274 may
be another source where an outburst pair has been detected where
the second giant outburst is less luminous than the first outburst.

The gap between the 1998-2001, 2010-2011, 2018 and
2021 outbursts in XTE J1946+274 are $\sim$19.2 $\rm{P_{orb}}$, $\sim$14.9 $\rm{P_{orb}}$ and $\sim$6.4 $\rm{P_{orb}}$ (using $\rm{P_{orb}}$ $\sim$171 d). This inferred time-scale is similar to the estimated time-scale
between two giant outbursts (about $\sim$12 $\rm{P_{orb}}$) obtained from numerical hydrodynamical simulations of ZLK
oscillations in the Be disc and the accretion disc around the neutron
star (\citealt{martin2021kozai}). Similar time-scales between consecutive giant outbursts have been observed in other Be/XRB sources such as in 4U 0115+63, 1A 0535+262 and SAX J2103.5+4545. Although the gap between the 2018 and the 2021 outbursts in XTE J1946+274 is slightly less ($\sim$6.4 $\rm{P_{orb}}$, roughly half of 12 $\rm{P_{orb}}$) but interestingly, the 2021 outburst was fainter by about a factor of $\sim$2 as suggested by the hydrodynamical simulation by \citealt{martin2021kozai} for Be/XRBs. A detailed hydrodynamic simulation of this system is required to explore and better understand the ZLK disc oscillations mediated giant outbursts in this enigmatic Be/XRB which is beyond the scope of this paper.

\section{Summary and conclusion}
We have detected broad-band X-ray pulsations from XTE J1946+274 using the SXT and
LAXPC instruments onboard the \textit{AstroSat} mission during the rare 2018 and 2021 outbursts of this Be/XRB system. The 2018 and 2021 outbursts were single outbursts
unlike the prolonged outburst series detected in 1998 and 2010. We construct the spin evolution of the pulsar for over two decades and find that the pulsar spins up during outbursts and usually spins down during dormant periods. Broad-band energy resolved pulse profiles of the pulsar have been generated which evolve with energy and subtle changes in profiles during the two singular outbursts studied with \textit{AstroSat} are detected. The energy spectrum of the pulsar has been derived over the 0.5-80 keV band from
the combined SXT and LAXPC observations. We find inkling of reversal in the correlation between the cyclotron line energy and luminosity which needs to be verified from future observations of this pulsar. We discuss possible mechanisms which can drive these unusual Type II outbursts in this Be/XRB system and suggest that multiwavelength observations and detailed hydrodynamic simulations of this system can shed more light on the underlying phenomena which triggers rare motley of outbursts in this source.

\textit{Note: A very recent work \cite{devaraj2022phase} came to our notice after the preparation of our manuscript. A quick glance at the manuscript shows similar inference for the 2018 outburst and in any case our results for the 2021 outburst are new.}

\normalem
\begin{acknowledgements}
We are extremely thankful to the reviewer for carefully
going through the manuscript and making valuable and
constructive suggestions that have improved the presentation of this paper.
This publication uses the data from the \textit{AstroSat} mission of the Indian Space
Research Organisation (ISRO), archived at the Indian Space Science
Data Centre (ISSDC). We thank members of LAXPC instrument
team at TIFR and the \textit{AstroSat} project team at URSC for their
contributions to the development of the LAXPC instrument. We
thank the LAXPC POC at TIFR for verifying and releasing the
data. \texttt{LAXPCSOFT} software (\url{http://astrosat-ssc.iucaa.in/laxpcData}) is
used for analysis in this paper. This work has
also used the data from the Soft X-ray Telescope (SXT) developed at
TIFR, Mumbai, and the SXT POC at TIFR is thanked for verifying
and releasing the data via the ISSDC data archive and providing the
necessary software tools. This research has made use of software provided by the High Energy Astrophysics Science Archive Research Center (HEASARC), which is a service of the Astrophysics Science Division at NASA/GSFC and the High Energy Astrophysics Division
of the Smithsonian Astrophysical Observatory. This research has also made use of
the \textit{Fermi}/GBM (\citealt{meegan2009fermi}) pulsar spin evolution history
provided by the \textit{Fermi} team. This research has also made use of the \textit{CGRO}/BATSE pulsar spin evolution history provided by the BATSE team. This research has made use of the \textit{MAXI} light curve provided by RIKEN, JAXA, and the \textit{MAXI} team.
\end{acknowledgements}

\section*{Data availability}
This research has made use of data from the \textit{AstroSat} mission archived at the Indian Space Science Data Centre (ISSDC). The archival data from the \textit{AstroSat} mission is publicly available at
\url{https://www.issdc.gov.in/astro.html}. The \textit{CGRO}/BATSE pulsar spin evolution history for XTE J1946+274 provided by the BATSE team is available at\\ \url{https://gammaray.nsstc.nasa.gov/batse/pulsar/data/sources/groj1944.html}. The \textit{Fermi}/GBM pulsar spin evolution history for XTE J1946+274 provided by the \textit{Fermi} team is available at\\ \url{https://gammaray.nsstc.nasa.gov/gbm/science/pulsars/lightcurves/xtej1946.html}.

\bibliographystyle{raa}
\bibliography{main.bib}

\end{document}